\documentclass[floatfix,prb,twocolumn,showpacs,superscriptaddress,notitlepage]{revtex4-1}

\usepackage{amsmath}
\usepackage{graphicx}
\usepackage{amssymb}
\usepackage{dcolumn}
\usepackage{bm}
\usepackage{longtable}
\usepackage{color}

\usepackage[T1]{fontenc}
\usepackage[utf8]{inputenc}
\usepackage{textcomp}
\usepackage{txfonts}
\usepackage{multirow}
\usepackage{hyperref}

\DeclareMathAlphabet{\mathitbf}{T1}{cmr}{bx}{it}

\newcommand{\Tc}{T_\mathrm{c}}

\newcommand{\vn}[1]{\boldsymbol{ #1}}

\begin{document}

\title{Critical parameters of the three-dimensional Ising spin glass}

\author{M. Baity-Jesi} \affiliation{Departamento
  de F\'\i{}sica Te\'orica I, Universidad
  Complutense, 28040 Madrid, Spain.} \affiliation{Dipartimento di Fisica, La Sapienza Universit\`a di Roma,
     00185 Roma,  Italy.}\affiliation{Instituto de Biocomputaci\'on y
  F\'{\i}sica de Sistemas Complejos (BIFI), 50009 Zaragoza, Spain.}

\author{R.~A.~Ba\~nos} \affiliation{Instituto de Biocomputaci\'on y
  F\'{\i}sica de Sistemas Complejos (BIFI), 50009 Zaragoza, Spain.}
  \affiliation{Departamento
  de F\'\i{}sica Te\'orica, Universidad
  de Zaragoza, 50009 Zaragoza, Spain.} 

\author{A.~Cruz} \affiliation{Departamento
  de F\'\i{}sica Te\'orica, Universidad
  de Zaragoza, 50009 Zaragoza, Spain.} \affiliation{Instituto de Biocomputaci\'on y
  F\'{\i}sica de Sistemas Complejos (BIFI), 50009 Zaragoza, Spain.}

\author{L.A.~Fernandez} \affiliation{Departamento
  de F\'\i{}sica Te\'orica I, Universidad
  Complutense, 28040 Madrid, Spain.} \affiliation{Instituto de Biocomputaci\'on y
  F\'{\i}sica de Sistemas Complejos (BIFI), 50009 Zaragoza, Spain.}

\author{J.~M.~Gil-Narvion} \affiliation{Instituto de Biocomputaci\'on y
  F\'{\i}sica de Sistemas Complejos (BIFI), 50009 Zaragoza, Spain.}

\author{A.~Gordillo-Guerrero}\affiliation{D. de  Ingenier\'{\i}a
El\'ectrica, Electr\'onica y Autom\'atica, U. de Extremadura,
  10071, C\'aceres, Spain.}\affiliation{Instituto de Biocomputaci\'on y
  F\'{\i}sica de Sistemas Complejos (BIFI), 50009 Zaragoza, Spain.}

\author{D.~I\~niguez} \affiliation{Instituto de Biocomputaci\'on y F\'{\i}sica
  de Sistemas Complejos (BIFI), 50009 Zaragoza, Spain.}
\affiliation{Fundaci\'on ARAID, Diputaci\'on General de Arag\'on, Zaragoza, Spain.}

\author{A.~Maiorano} \affiliation{Dipartimento di Fisica, La Sapienza Universit\`a di Roma,
     00185 Roma,  Italy.}\affiliation{Instituto de Biocomputaci\'on y F\'{\i}sica de Sistemas
  Complejos (BIFI), 50009 Zaragoza, Spain.}

\author{F.~Mantovani} \affiliation{Dipartimento di Fisica e Scienze della
  Terra, Universit\`a di Ferrara, and INFN, Ferrara, Italy.}

\author{E.~Marinari}\affiliation{Dipartimento di Fisica, IPCF-CNR, UOS
Roma Kerberos and INFN, La Sapienza Universit\`a di Roma, 00185 Roma,  Italy.} 

\author{V.~Martin-Mayor} \affiliation{Departamento de F\'\i{}sica Te\'orica I,
  Universidad Complutense, 28040 Madrid, Spain.} \affiliation{Instituto de
  Biocomputaci\'on y F\'{\i}sica de Sistemas Complejos (BIFI), 50009 Zaragoza,
  Spain.}

\author{J.~Monforte-Garcia} \affiliation{Instituto de Biocomputaci\'on y
  F\'{\i}sica de Sistemas Complejos (BIFI), 50009 Zaragoza, Spain.}
  \affiliation{Departamento
  de F\'\i{}sica Te\'orica, Universidad
  de Zaragoza, 50009 Zaragoza, Spain.} 

\author{A.~Mu\~noz Sudupe} \affiliation{Departamento
  de F\'\i{}sica Te\'orica I, Universidad
  Complutense, 28040 Madrid, Spain.} 

\author{D.~Navarro} \affiliation{D.  de Ingenier\'{\i}a,
  Electr\'onica y Comunicaciones and I3A, U. de Zaragoza, 50018 Zaragoza, Spain.}

\author{G.~Parisi}\affiliation{Dipartimento di Fisica, IPCF-CNR, UOS
Roma Kerberos and INFN, La Sapienza Universit\`a di Roma, 00185 Roma,  Italy.}

\author{S.~Perez-Gaviro} \affiliation{Instituto de Biocomputaci\'on y
  F\'{\i}sica de Sistemas Complejos (BIFI), 50009 Zaragoza, Spain.}
\affiliation{Fundaci\'on ARAID, Diputaci\'on General de Arag\'on, Zaragoza, Spain.}

\author{M.~Pivanti} \affiliation{Dipartimento di Fisica e Scienze della
  Terra, Universit\`a di Ferrara, and INFN, Ferrara, Italy.} 

\author{F.~Ricci-Tersenghi}\affiliation{Dipartimento di Fisica, IPCF-CNR, UOS
Roma Kerberos and INFN, La Sapienza Universit\`a di Roma, 00185 Roma,  Italy.}  

\author{J.~J.~Ruiz-Lorenzo} \affiliation{Departamento de F\'{\i}sica,
  Universidad de Extremadura, 06071 Badajoz, Spain.}\affiliation{Instituto de
  Biocomputaci\'on y F\'{\i}sica de Sistemas Complejos (BIFI), 50009 Zaragoza,
  Spain.}

\author{S.F.~Schifano} \affiliation{Dipartimento di Matematica e Informatica, Universit\`a di Ferrara
and INFN, Ferrara, Italy}

\author{B.~Seoane}\affiliation{Dipartimento di Fisica, La Sapienza Universit\`a di Roma,
     00185 Roma,  Italy.} \affiliation{Instituto de
  Biocomputaci\'on y F\'{\i}sica de Sistemas Complejos (BIFI), 50009 Zaragoza,
  Spain.}

\author{A.~Tarancon} \affiliation{Departamento
  de F\'\i{}sica Te\'orica, Universidad
  de Zaragoza, 50009 Zaragoza, Spain.} \affiliation{Instituto de Biocomputaci\'on y
  F\'{\i}sica de Sistemas Complejos (BIFI), 50009 Zaragoza, Spain.}

\author{R.~Tripiccione} \affiliation{Dipartimento di Fisica e Scienze della
  Terra, Universit\`a di Ferrara, and INFN, Ferrara, Italy.} 
 
\author{D.~Yllanes}\affiliation{Dipartimento di Fisica, La Sapienza Universit\`a di Roma,
     00185 Roma,  Italy.}  \affiliation{Instituto de
  Biocomputaci\'on y F\'{\i}sica de Sistemas Complejos (BIFI), 50009 Zaragoza,
  Spain.}

\begin{abstract}
We report a high-precision finite-size scaling study of the critical behavior
of the three-dimensional Ising Edwards-Anderson model (the Ising spin
glass). We have thermalized lattices up to $L=40$ using the Janus dedicated
computer. Our analysis takes into account leading-order corrections to
scaling.  We obtain $T_\mathrm{c}=1.1019(29)$ for the critical temperature,
$\nu=2.562(42)$ for the thermal exponent, $\eta=-0.3900(36)$ for the anomalous
dimension and $\omega=1.12(10)$ for the exponent of the leading corrections to
scaling. Standard (hyper)scaling relations yield $\alpha= -5.69(13)$,
$\beta=0.782(10)$ and $\gamma = 6.13(11)$.  We also compute several
universal quantities at $T_\mathrm{c}$.
\end{abstract}

\pacs{75.50.Lk, 75.40.Mg} 
\maketitle 

\section{Introduction}\label{sec:intro}
Spin glasses are disordered magnetic alloys whose understanding has defied
physicists for decades.\cite{mydosh:93,mezard:87} In this context, the Ising
Edwards-Anderson model\cite{edwards:75} has played a major role. However, in
spite of its prominence, it took 25 years to show that it undergoes a
continuous phase transition at a critical temperature
$T_\mathrm{c}$\cite{palassini:99,ballesteros:00} (there was an earlier
consensus on the existence of a phase
transition,\cite{kawashima:96,iniguez:96,marinari:98d,iniguez:97,berg:98,janke:98b}
but its nature had remained unclear). Amusingly, evidence for a phase
transition on experimental spin glasses had been obtained several years
before.\cite{gunnarsson:91}

Since then, the critical behavior of the Edwards-Anderson model has been
studied numerically in a number of
papers.\cite{mari:02,nakamura:03,daboul:04,pleimling:05,perez-gaviro:06,parisen:06,jorg:06,campbell:06,katzgraber:06,machta:08,hasenbusch:08,hasenbusch:08b}
In these works, microscopic details such as the distribution of the coupling
constants differ. It was unclear whether universality violations were present
in the problem because the critical exponents and other universal quantities
seemed to depend on those microscopic details (although some
authors\cite{katzgraber:06,jorg:06,jorg:08c} argued that these apparent
violations were caused by corrections to scaling). The issue was settled in
2008 by Hasenbusch, Pelissetto and Vicari,\cite{hasenbusch:08b} who emphasized
the role of corrections to scaling, thus convincing the community that
universality holds.  Furthermore, their computation of most universal
quantities is still the most accurate to date.

Here we present a high-precision finite-size scaling study of the Ising
Edwards-Anderson model. Using the Janus special-purpose
computer,\cite{janus:08,janus:09} we thermalize the largest lattices to date
($L=40$), with a very large number of samples. Even with this increased
accuracy, we confirm that the analysis with leading-order scaling corrections
is adequate (however, see below Sect.~\ref{sect:dimensionless}). In this way,
we achieve a determination of the critical exponents four times more accurate
than the one in Ref.~\onlinecite{hasenbusch:08b}.  We also compute a number of
universal quantities not previously considered in the literature.  Reliable
determinations of the critical parameters are important to make progress in
other fronts, such as the study of the correlation functions below
$\Tc$,\cite{janus:10b} or the behavior of spin glasses in an externally
applied magnetic field.\cite{janus:12,janus:13}

The organization of the remaining part of this work is as follows. In
Section~\ref{sect:model-simulations} we define the model and provide details
about our simulations. The quantities that we compute are defined in
Section~\ref{sect:observables}.  The finite-size scaling analysis is briefly
reviewed in Section~\ref{sect:FSS}.  Our main results are given in
Section~\ref{sect:mega-fit}, where we compute the critical exponents,
including the corrections to scaling exponent $\omega$ (see also Appendix
\ref{sec:omega-alt}), as well as the critical correlation length in units of
the lattice size $\xi_L/L$ and the Binder cumulant $U_4$. With this input, we
proceed to compute in Section~\ref{sect:other} other universal cumulants and
the critical temperature. Finally, we discuss our
conclusions in Section~\ref{sect:conclusions}. For ease of reference our main
results are summarized in Table~\ref{tab:parameters}.
\begin{table}[t]
\begin{ruledtabular}
\begin{tabular}{D{=}{\,\ =\,\ }{7}c}
\multicolumn{1}{c}{$\qquad\quad$ Quantity}  &Source  \\
\hline
 \omega =1.12(10)   & \multirow{5}{*}{Joint fit}\\
\eta = -0.3900(36) \\
\nu = 2.562(42)    \\
R_\xi^* = 0.6516(32)   \\
U_4^* = 1.4899(28)    \\
\hline
\alpha =  -5.69(13) &\multirow{3}{*}{$\qquad$ Derived quantities $\qquad$}\\
\beta = 0.782(10) & \\
\gamma = 6.13(11)  & \\
\hline
T_{c} = 1.1019(29) &   \multirow{6}{*}{Secondary fits}\\
\qquad U_{1111}^* =  0.4714(14) & \\
U_{22}^* = 0.7681(16) &  \\
U_{111}^* = 0.4489(15) & \\
B_\chi^* =   2.4142(51) &\\
R_{12}^* = 2.211\,\pm\, 0.006  &  \\
\end{tabular}
\end{ruledtabular}
\caption{Summary of our results for the universality class of the Ising spin
  glass (see definitions in Sect.~\ref{sect:observables}).  The first block of
  five quantities comes from a joint fit reported in Fig.~\ref{fig:conjunto}
  and Sect.~\ref{sect:mega-fit}. The second block of quantities
  includes the remaining critical exponents, which can be derived from $\nu$
  and $\eta$ (taking correlations into account for the errors).
  The third block of quantities come from secondary individual fits. 
  The computation of the critical temperature
  $T_\text{c}$ is reported in  Fig.~\ref{fig:betac} and Sect.~\ref{sect:Tc}.
  Finally, the remaining universal quantities are computed in
  Sect.~\ref{sect:dimensionless}.  In all cases we have employed the
  quotients method and performed fits with leading corrections to scaling for
  all data with $L\geq L_\text{min}=8$. Since we computed all the covariance
  matrices from $\mathcal O(10^3)$ jackknife blocks, our error estimates are
  significant beyond the first digit. The error for $R_{12}^*$ is of a
  systematic nature (rather than statistical, see
  Sect.~\ref{sect:dimensionless}).
\label{tab:parameters}}
\end{table}

\section{Model, simulations, observables}\label{sect:model}

\subsection{Model and simulations}\label{sect:model-simulations}

We consider Ising spins $s_{\vn{x}}=\pm 1$, defined on the $V=L^D$ nodes of a
cubic lattice of linear size $L$ and spatial dimension $D=3$, with periodic
boundary conditions. The interactions in the Hamiltonian $H$ are restricted to
lattice nearest neighbors:
\begin{equation}\label{eq:H-definition}
H=-\sum_{\langle {\vn{x},\vn{y}} \rangle} J_{\vn{x},\vn{y}} s_{\vn{x}}s_{\vn{y}}\,.
\end{equation}
The coupling constants $J_{\vn{x},\vn{y}}$ can take the two values $\pm1$ with
$50\%$ probability. We study quenched disorder, meaning that the
$J_{\vn{x},\vn{y}}$ cannot change with time (see, e.g., Ref.~\onlinecite{mezard:87}).
Each instance of the $\{J_{\vn{x},\vn{y}}\}$ is called \emph{sample}.  For any
quantity of interest $O$, we first compute the thermal average $\langle O
\rangle$ and only afterwards we take the average over the different samples
$\overline{\langle O \rangle}$.

For every sample we  simulate four real replicas $\{s_{\vn{x}}^a\}$, $a=1,2,3$
and $4$. All four replicas share the same set of coupling constants
$\{J_{\vn{x},\vn{y}}\}$, but they are otherwise statistically independent.

We employ parallel tempering.\cite{hukushima:96,marinari:98b} We simulate
lattices of size up to $L=24$ on the \emph{Memento} CPU cluster at BIFI.
Multi-spin coding with streaming extensions allows us to simulate 128 samples
in parallel. On the other hand, lattices of linear sizes $L=32$ and $40$ are
simulated on Janus. The main features of our simulations are reported in
Table~\ref{tab:simulations}.  As a whole, the simulations on \emph{Memento} for
lattice sizes $L\leq 24$ implied a total of $2.99\times10^{19}$ Metropolis
spin updates (the equivalent of $1.33\times10^5$ days of a single core of the
machine). The simulations on Janus ($L=32,40$) consisted of a total of
$5.03\times10^{19}$ heat-bath spin updates, equivalent to about $27\,400$ days
of a single processing unit (FPGA).

We have checked that our data are not affected by thermalization effects.  For
the largest lattice sizes ($L=32,40$) we use the method reported in
Ref.~\onlinecite{janus:10}, which consists in computing the exponential
autocorrelation time $\tau_\text{exp}$ for each sample, using the temperature
random walk during the parallel tempering. We extend each sample until the
simulation time is at least $16\tau_\text{exp}$ (therefore, the length of the
simulation depends on the sample, as shown on Table~\ref{tab:simulations}).
For the lattices simulated with multi-spin coding, this sample-by-sample
method is more involved.\cite{janus:12} Therefore, taking into account that
almost all our observables are measured during the simulation, we have decided
to use the more traditional approach of studying the time evolution of
sample-averaged quantities on a logarithmic scale. All the quantities that we
have considered are stable on the last two logarithmic bins, corresponding to
the second half and the second quarter of the run. In fact, this condition is
satisfied even if we subtract from each successive bin the result over the
last half of the measurements, thus significantly reducing the error
bars.\cite{fernandez:08b}

In general, we compute all the physical quantities by averaging over
the second half of the simulation. However, as a further check, we have
also recomputed all the final quantities using only the last block 
of measurements (which, depending on the lattice, corresponds from
$6\%$ to $25\%$ of the total simulation time). We find no differences
greater than one fifth of a standard deviation (which, in any case, 
corresponds to the increase in the statistical error of thermal averages).

\begin{table}
\begin{ruledtabular}
\begin{tabular}{ccccccc}
$L$ & $N_\mathrm{samples}$ & $N_\mathrm{MCS}^{\mathrm{min}}$& $N_\mathrm{MCS}^{\mathrm{max}}$& $N_\mathrm{T}$  & $T_\mathrm{min}$ & $T_\mathrm{max}$\\\hline\hline
  6   & 8\,192\,000 &     40\,000  & 40\,000        & 10 & 1.100 & 1.703\\
  8   & 8\,192\,000 &     80\,000  & 80\,000        & 10 & 1.100 & 1.703\\
  10  & 8\,192\,000 &     80\,000  & 80\,000        & 10 & 1.100 & 1.703\\
  12  & 8\,192\,000 &     80\,000  & 80\,000        & 14 & 1.100 & 1.651\\
  16  & 1\,024\,000 &    800\,000  & 800\,000       & 14 & 1.100 & 1.651\\  
  20  &    768\,000 & 1\,600\,000  & 1\,600\,000    & 14 & 1.100 & 1.651\\
  24  &    512\,000 & 3\,200\,000  & 3\,200\,000    & 23 & 1.100 & 1.626\\
  32  &    256\,000 & 1\,600\,000  &  99\,200\,000  & 22 & 1.100 & 1.600\\
  40  &     48\,000 & 6\,400\,000  & 204\,800\,000  & 28 & 1.100 & 1.594\\
\end{tabular}
\end{ruledtabular}
 \caption{Details of the simulations. We show the simulation parameters for
   each lattice size $L$.  $N_\mathrm{samples}$ is the number of simulated
   samples.  $N_\mathrm{T}$ is the number of temperatures that were used in
   parallel tempering. In the set of temperatures we always include the values
   1.1, 1.11266, 1.12532, 1.13797, and evenly space the remaining
   $N_\mathrm{T}-4$ temperatures up to $T_\mathrm{max}$ (the temperature
   resolution was increased near $\Tc$ in order to ease interpolations).  The
   number of temperatures $N_\mathrm{T}$ was chosen so that the parallel
   tempering's acceptance was at least of $15\%$.
   $N_\mathrm{MCS}^{\mathrm{min}}$ is the minimum number of Monte Carlo steps
   (MCS) in each simulation.  Each MCS consisted of 10 Metropolis (heat-bath
   in $L=32,40$) full-lattice sweeps, followed by a parallel-tempering
   temperature swap. In the larger lattices ($L=32$, $40$) we extend the
   simulation of specific samples after measuring the exponential correlation
   time.\cite{janus:10} The average simulation time was larger than the
   minimal one by a factor 1.6 ($L=32$) or 1.4
   ($L=40$).} \label{tab:simulations}
\end{table}

\subsection{Observables}\label{sect:observables}

The main quantities are computed in terms of the overlap field
\begin{equation}
q_{\vn{x}}^{ab}=s_{\vn{x}}^a s_{\vn{x}}^b\,.
\end{equation}
Its spatial correlation function is
\begin{equation}
G(\vn{r})=\frac{1}{V}\sum_{\vn{x}} \overline{\langle q_{\vn{x}+\vn{r}}^{ab} q_{\vn{x}}^{ab}\rangle}\,,
\end{equation}
while the spin-glass order parameter is the spatial average
\begin{equation}
q^{ab}=\frac{1}{V}\sum_{\vn{x}}  q_{\vn{x}}^{ab}\,.
\end{equation}
The reader will notice that, having four replicas at our disposal, there are
six equivalent ways of choosing the pair of replica indices $ab$. We shall
merely write $q$ to imply that we average over all possible replica index
combinations in order to improve our statistics.

The second-moment correlation length is computed from the Fourier transform
of the correlation function
\begin{equation}
\chi(\vn{k})=\frac{1}{V}\sum_{\vn{r}}G(\vn{r})\,\mathrm{e}^{\mathrm{i} \vn{k}\cdot\vn{r}}\,.
\end{equation}
Specifically,\cite{cooper:82,amit:05} 
\begin{equation}\label{eq:xi-second-moment}
\xi = \frac{1}{2 \sin (k_\mathrm{min}/2)} \sqrt{\frac{\chi(0)}{\chi(\vn{k}_\mathrm{min})} -1},
\end{equation}
where ${\vn{k}_\mathrm{min}}=(2\pi/L,0,0)$ or permutations.  We remark as
well that the spin-glass susceptibility is
\begin{equation}
\chi_\mathrm{SG}= \chi(0) = V \overline{\langle q^2\rangle}\,.
\end{equation}
We shall often study the correlation length in units of the system size
\begin{equation}\label{eq:Rxi}
R_\xi = \xi/L,
\end{equation}
whose value is universal at the critical point.

It will be useful to consider six more dimensionless quantities, which are
also universal at $\Tc$:
\begin{align}
U_4&=\frac{\overline{\langle q^4\rangle}}{\overline{\langle
    q^2\rangle}^2}\,,\label{eq:U4-def}\\[1.3ex]
U_{22}&=\frac{\overline{\langle q^2\rangle^2}}{\overline{\langle
    q^4\rangle}}\,,\label{eq:U22-def}\\[1.3ex]
U_{111}&=\frac{\overline{\langle
    q^{12}q^{23}q^{31}\rangle}^{\,4/3}}{\overline{\langle q^4\rangle}}\,,\label{eq:U111-def}\\[1.3ex]
U_{1111}&=\frac{\overline{\langle q^{12}q^{23}q^{34} q^{41}\rangle}}{\overline{\langle q^4\rangle}}\,,\label{eq:U1111-def}\\[1.3ex]
R_{12}&=\frac{\chi(2\pi/L,0,0)}{\chi(2\pi/L,2\pi/L,0)},\label{eq:R12-def}\\[1.3ex]
B_\chi &= 3 V^2 \frac{\overline{\langle |\hat q
    (2\pi/L,0,0)|^4\rangle}}{[\chi(2\pi/L,0,0)]^2}\,,\label{eq:Bchi-def}
\end{align}
where
\begin{equation}
\hat q^{ab}(\vn{k}) = \frac{1}{V} \sum_{\vn{x}}\ q^{ab}_{\vn{x}}\,
\mathrm{e}^{\mathrm{i} \vn{k}\cdot\vn{x}}\,.
\end{equation}
In order to gain statistics, we average over all equivalent wave-vectors in
Eqs.~\eqref{eq:R12-def} and~\eqref{eq:Bchi-def}. Similarly, we average over
all the equivalent choices for the replica indices in Eqs.~\eqref{eq:U111-def}
and~\eqref{eq:U1111-def}. We recall that $R_{12}$ was crucial to understand
the critical behavior in a magnetic field.\cite{janus:12} Some of the other 
quantities have been studied before.\cite{billoire:11}

Temperature derivatives are computed in two ways. We either use the connected
correlations with the energy, or we perform a third-order polynomial
interpolation and differentiate it. We have found that both determinations
differ only in a small fraction of the error bars (which were computed using
the jackknife method, see, e.g., Ref.~\onlinecite{amit:05}). In our final
results, we have employed the interpolation-polynomial method.
\begin{figure}[t]
\includegraphics[height=\linewidth, angle=270]{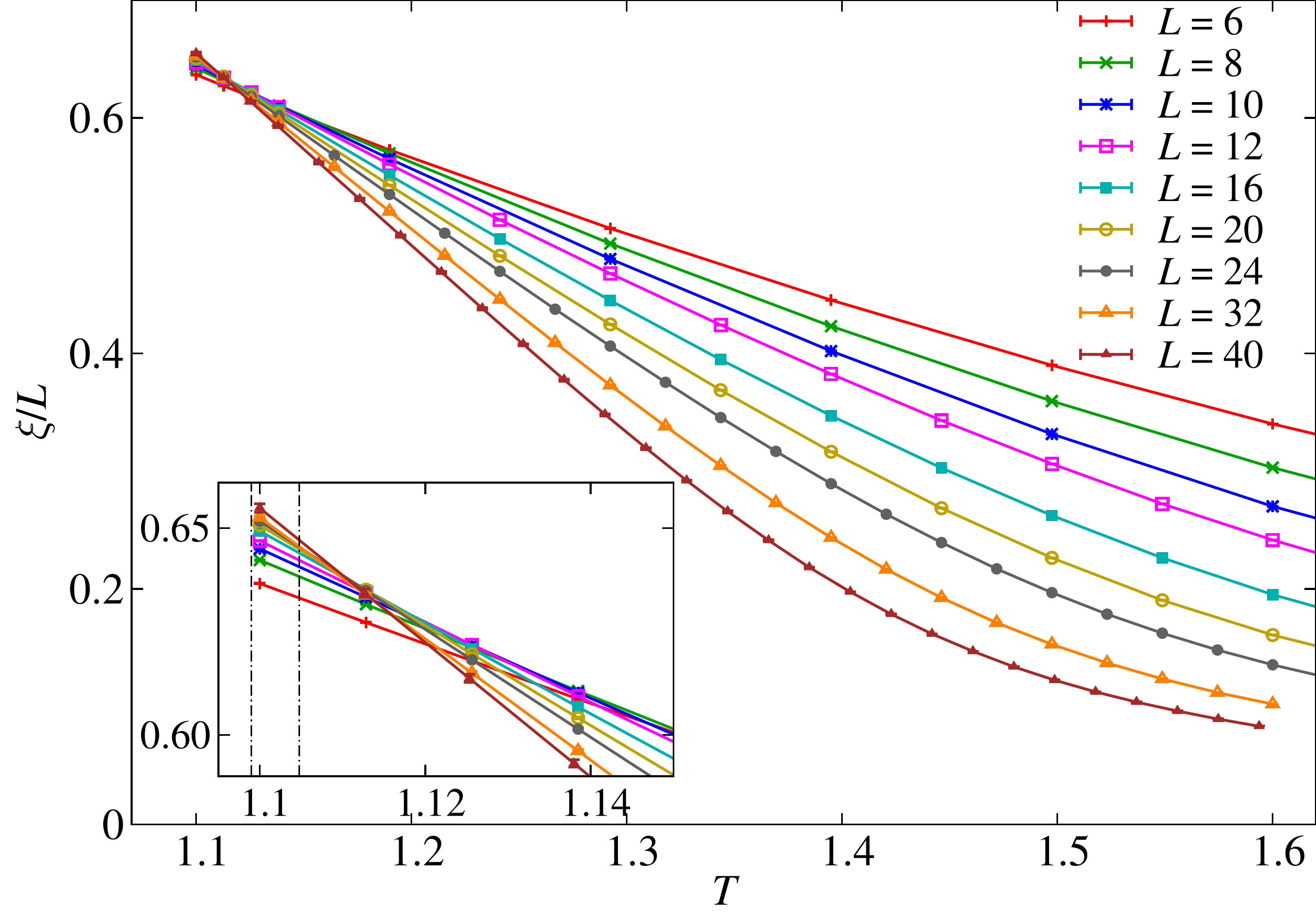} \caption{(color online).
  Plot of the second-moment correlation length $\xi$,
  Eq.~\eqref{eq:xi-second-moment}, in units of the lattice size $L$ for all
  our simulated systems as a function of temperature. The inset is a detailed
  view of the critical region, showing scale invariance, where the vertical
  lines mark our final estimate for (the error interval of) the critical
  temperature, $T_\text{c} = 1.1019(29)$.\label{fig:cortes}}
\end{figure}

\section{Finite-Size Scaling analysis}\label{sect:FSS}
To extract the value of critical points, critical exponents and dimensionless
quantities, we employ the quotients
method,\cite{nightingale:76,ballesteros:96,amit:05} also known as
phenomenological renormalization. This method allows a particularly transparent
study of corrections to scaling. Previous applications to disordered systems
include diluted ferromagnets,\cite{ballesteros:98b} spin glasses\cite{ballesteros:00,campos:06,jorg:06,jorg:08c,fernandez:09b,leuzzi:08,banos:12,baityjesi:13}
and systems belonging to the random-field Ising model
realm.\cite{fernandez:11b,fernandez:12,fytas:13}

The method is actually very simple.  We compare observables computed in pairs
of lattices $(L,2L)$. We start by imposing scale invariance. We look for the
$L$-dependent critical point: the value of $T$ such that $\xi_{2L}/\xi_L=2$
(i.e., the \emph{crossing} point for $R_\xi=\xi_L/L$, see
Fig~\ref{fig:cortes}).

Now, for dimensionful quantities $O$, which scale as
$\xi^{x_O/\nu}$ in the thermodynamical limit, we consider the quotient
${\mathcal Q}_O=O_{2L}/O_L$ at the crossing. Instead, for dimensionless
quantities $g$ the ratio $g_{2L}/g_L$ trivially goes to one, therefore we
focus on $g_{L}$.  In either case, one has:
\begin{equation}\label{eq:QO}
{\mathcal Q}_O^{\,\mathrm{cross}}=2^{x_O/\nu}+\mathcal{O}(L^{-\omega})\,,\
g_{L}^{\,\mathrm{cross}}=g^\ast+\mathcal{O}(L^{-\omega})\,,
\end{equation}
where $x_O/\nu$, $g^\ast$ and the scaling-corrections exponent $\omega$ are
universal.  Examples of dimensionless quantities are $R_\xi$, the six
cumulants defined in Eqs.~(\ref{eq:U4-def}--\ref{eq:R12-def}).  Instances of
dimensionful quantities are the temperature derivatives of $\xi$
($x_{\partial_T \xi}=1+\nu$), the temperature derivatives of each of the six
cumulants ($x_{\partial_T g}=1$), and the susceptibility $\chi$ [$x_\chi=
  \nu(2-\eta)$].

The reader may observe that studying $g_L$ rather than $g_{2L}$ in
Eq.~\eqref{eq:QO} is somehow arbitrary. In fact, the relative size of
scaling corrections cannot be decided a priori.\footnote{Dimensionless
quantities behave as a function of $L$ and the reduced temperature
$t=(T-\Tc)/\Tc$ as
$$g(L,t)=f_g(L^{1/\nu} t) + L^{-\omega} h_g(L^{1/\nu}t)+\ldots\,,$$ where
$f_g$ and $h_g$ are very smooth (actually analytical) scaling functions.  In
particular, we name $f_\xi$ and $g_\xi$ the scaling functions corresponding to
$\xi/L$.  In fact, see e.g. Ref.~\onlinecite{amit:05}, the detailed form of
Eq.~\eqref{eq:Tc-shift} follows from the Taylor expansions
$f_\xi(x)=f_\xi(0)+x f'_\xi(0)+\ldots$ and $h_\xi(x)=h_\xi(0)+\ldots$:
$$
t^{\,\mathrm{cross}}_{L,2L}=\frac{h_\xi(0)}{f'_\xi(0)}\,
\frac{1-2^{-\omega}}{2^{1/\nu}-1}\, L^{-\omega-\frac{1}{\nu}}\ +\ \ldots\,.
$$
A similar computation yields the amplitudes for the scaling corrections of $g_{2L}^\mathrm{cross}$ and $g_L^\mathrm{cross}$ in
Eq.~\eqref{eq:QO}:
$$A_g^{(2L)}=2^{1/\nu}\frac{1-2^{-\omega}}{2^{1/\nu}-1}h_\xi(0)\frac{f_g'(0)}{f_\xi'(0)}\ +\ 2^{-\omega}
h_g(0)\,,$$
and
$$A_g^{(L)}=\frac{1-2^{-\omega}}{2^{1/\nu}-1}h_\xi(0)\frac{f_g'(0)}{f_\xi'(0)}\ +\ h_g(0)\, .$$
Either of the two amplitudes $A_g^{(L)}$, $A_g^{(2L)}$ can dominate, depending
both on $g$ and on the magnitude chosen to find the crossing point ($\xi/L$,
$U_4$, etc.)} As a rule, we study $g_L$ because its statistical errors are
smaller. However, checking that this choice is immaterial will be an important
consistency check.

As a general rule, in this work we shall consider only the leading-order
corrections to scaling, $\mathcal{O}(L^{-\omega})$, that appear  in
Eq.~\eqref{eq:QO}. In some particular cases our statistical errors will be
small enough to resolve subleading corrections. We shall represent these
subleading corrections in an \emph{effective} way as a second-order polynomial
in $L^{-\omega}$. However, corrections of order $L^{-2\omega}$ are only a
subclass of the full set of subleading corrections (see, e.g.,
Ref.~\onlinecite{amit:05}).

As for  the crossing  temperature $T_\mathrm{c}^{(L,2L)}$,  we recall  that it
approaches $\Tc$ as
\begin{equation}\label{eq:Tc-shift}
T_\mathrm{c}^{(L,2L)}-\Tc=A L^{-(\omega+1/\nu)}+\ldots\,,
\end{equation}
where $A$ is a scaling amplitude and the dots stand for subleading
corrections.

Finally, we remark that $\xi/L$ in the above outlined analyses could be
replaced by any other of the six cumulants, such as for instance $U_{4}$.

\section{The critical exponents}\label{sect:mega-fit}
\begin{figure}[t]
\includegraphics[height=\linewidth, angle=270]{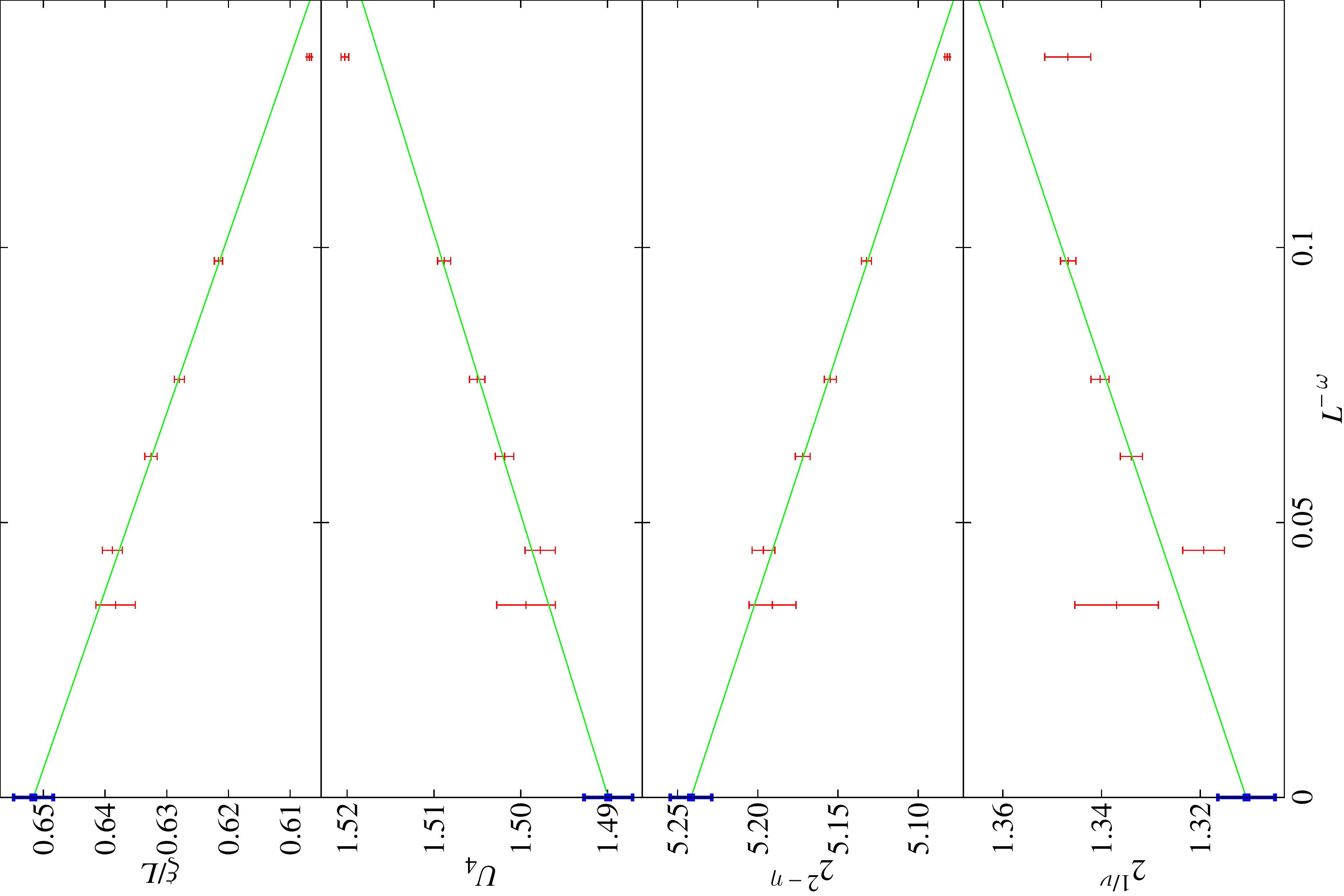}
\caption{(color online). Result of a joint fit yielding the critical exponents
  of the Ising spin glass. As discussed in the text, using the quotients
  method, we study the approach of two dimensionless universal quantities
  ($U_4$ and $\xi/L$) and of the quotients of two dimensionful quantities
  ($\chi$ and $\partial_T \xi$) to their critical value.  The rightmost points
  (corresponding to the crossings between $L=6$ and $L=12$) are not included
  in the fit.  The extrapolated values, reported in
  Table~\ref{tab:parameters}, are represented by a thick blue point on the $Y$
  axis. The fit, where the same $\omega=1.12(10)$ is used by the four
  quantities, has a $\chi^2/\text{d.o.f.} = 13.78/11$. The critical exponents
  are $\nu=2.562(42)$, $\eta=-0.3900(36)$.
\label{fig:conjunto}}
\end{figure}

Following the quotients method described in the previous section, we could
compute all the critical parameters (the critical exponent and the universal
values of dimensionless quantities) using fits to~\eqref{eq:QO}. For instance,
in order to compute the anomalous dimension $\eta$ we could use the relation
\begin{equation} 
\mathcal Q^\text{cross}_\chi = 2^{2-\eta} + A_\chi L^{-\omega}+\ldots\,.
\end{equation}
Hereafter, the dots will stand for subleading corrections to scaling.

In practice, of course, determining both the extrapolated value
and the value of the exponent $\omega$ in the same fit is very delicate, given
the low number of degrees of freedom available. In fact, the usual approach in
recent finite-size scaling studies has been to compute $\omega$ first,
using the behavior of dimensionless quantities, and then use this precomputed
value of $\omega$ to extrapolate the other critical exponents (see, e.g.,
Refs.~\onlinecite{janus:12,fytas:13} and also Appendix~\ref{sec:omega-alt}).
This approach has the disadvantage that all quantities have to be reported 
with two error bars (the first due to the statistical errors in the fit and the second 
due to the uncertainty in the precomputed $\omega$). Moreover, it does 
not take full advantage of the information contained in the critical behavior
of quotients of dimensionful quantities (because these are not used to 
refine the estimate of $\omega$).

In this paper, on the other hand, we consider all the most important 
quantities at the same time in a global fit. In particular, we take
as fitting functions
\begin{align}
U_4^\text{cross}(L) &= U_4^* + A_{U_4} L^{-\omega},\label{eq:U4}\\
R_\xi^\text{cross}(L) &= R_\xi^* + A_{\xi} L^{-\omega},\\
\mathcal Q^\text{cross}_\chi(L) &= 2^{2-\eta} + A_\chi L^{-\omega},\label{eq:Qchi}\\
\mathcal Q^\text{cross}_{\partial_T \xi/L}(L)  &= 2^{1/\nu}  + A_{\partial \xi} L^{-\omega}.\label{eq:QDxi}
\end{align}
Notice that $\omega$ is a common parameter in all of these functions. Then, we construct 
the $\chi^2$ goodness-of-fit estimator as 
\begin{equation}\label{eq:chi2-def}
\chi^2 = \sum_{i,j,a,b} \bigl[ y_i(L_a) - y_i^* -A_iL_a^{-\omega}\bigr]
[\sigma^{-1}]_{(ia)(jb)} \bigl[ y_j(L_b) -y_j^* - A_jL_b^{-\omega}\bigr],
\end{equation} 
where $a,b$ run over the system sizes, $L_a$ denotes the smaller $L$ in each
of the crossings $(L,2L)$, and $y_i$ is any of the $\mathcal Q^\text{cross}_O$
or of the $g^\text{cross}$ of Eqs.~\eqref{eq:U4}--\eqref{eq:QDxi}. The matrix
$\sigma^{-1}$ is the inverse of the full covariance matrix of the data.
This approach is statistically reliable and allows us to extract a large
amount of information from the numerical data.

We have plotted this joint fit in Figure~\ref{fig:conjunto}. We have discarded
the data from the $(L,2L)=(6,12)$ crossing, which clearly shows subleading corrections to 
scaling. The resulting fit, with $\chi^2/\text{d.o.f.} = 13.78/11$ ($P=25\%$) yields
the following critical parameters, defining the universality class of the Ising spin glass:
\begin{align} \label{eq:critical-parameters}
\omega &= 1.12(10), & 
\eta   &= -0.3900(36), &
\nu   &=  2.562(42),
\end{align}
\begin{align}\label{eq:universal-quantities}
R_\xi^* &= 0.6516(32), &
U_4^*  &= 1.4899(28).
\end{align}  
The amplitudes in the fit are
\begin{equation}
\begin{aligned}
A_\xi &= -0.309(42), & A_{U_4} &= 0.196(32), \\
A_\chi &= -0.141(20), & A_{\partial \xi} &= 0.374(70).
\end{aligned}
\end{equation}
 In addition, using the scaling and hyperscaling relations, we can
give the value of the remaining critical exponents (taking correlations into
account for the errors):
\begin{align}
\gamma &= 6.13(11), & \beta&=0.782(10), & \alpha&= -5.69(13).
\end{align}
More generally, for future reference, we report some correlation coefficients
(useful to compute the error in derived quantities)
\begin{align}
r_{\omega\nu} &= -0.58, & r_{\omega\eta} &= 0.75, & r_{\nu\eta} &=  -0.76,
\end{align}
where
\begin{equation}
r_{AB} = \frac{\text{Cov}(A,B)}{\sqrt{\text{Var}(A)\text{Var}(B)}}\ .
\end{equation} 

We remark that, in principle, we could have added the other
dimensionless quantities defined in Section~\ref{sect:observables} to the fit,
thus obtaining their values at the critical point as well as presumably
improving our determination of $\omega$. The problem, of course, 
is that there is only so much information in the system. If one keeps
adding quantities to the fit, eventually the covariance matrix becomes 
singular (or, at least, singular for numerical purposes). Therefore, 
in practice there is a limit to how many different quantities
can be analyzed at the same time.

Finally, we would like to mention that an alternative way of computing $\eta$
has been recently suggested.\cite{yllanes:11} One could compute the spin
correlation function in Fourier space, $\chi(\boldsymbol k)$, conditioned to a
fixed value of the spin overlap. In particular $\chi(\boldsymbol k)|_{q=0}$,
where all the thermal averages consider only those pairs of configurations
where $|q|$ is smaller than a certain window $q_0=\mathcal O(V^{-1/2})$. This
has the advantage of reducing the statistical errors significantly with
respect to the unrestricted correlation function. However, the fact that we
cannot use the $\boldsymbol k=0$ mode introduces stronger corrections to
scaling, so we have not followed this alternative approach to compute $\eta$
(but we have checked that it would give a consistent, though less accurate,
estimate).

\section{Other extrapolations to the thermodynamic limit}\label{sect:other}

\subsection{The critical temperature}\label{sect:Tc}
\begin{figure}[t]
\includegraphics[height=\linewidth, angle=270]{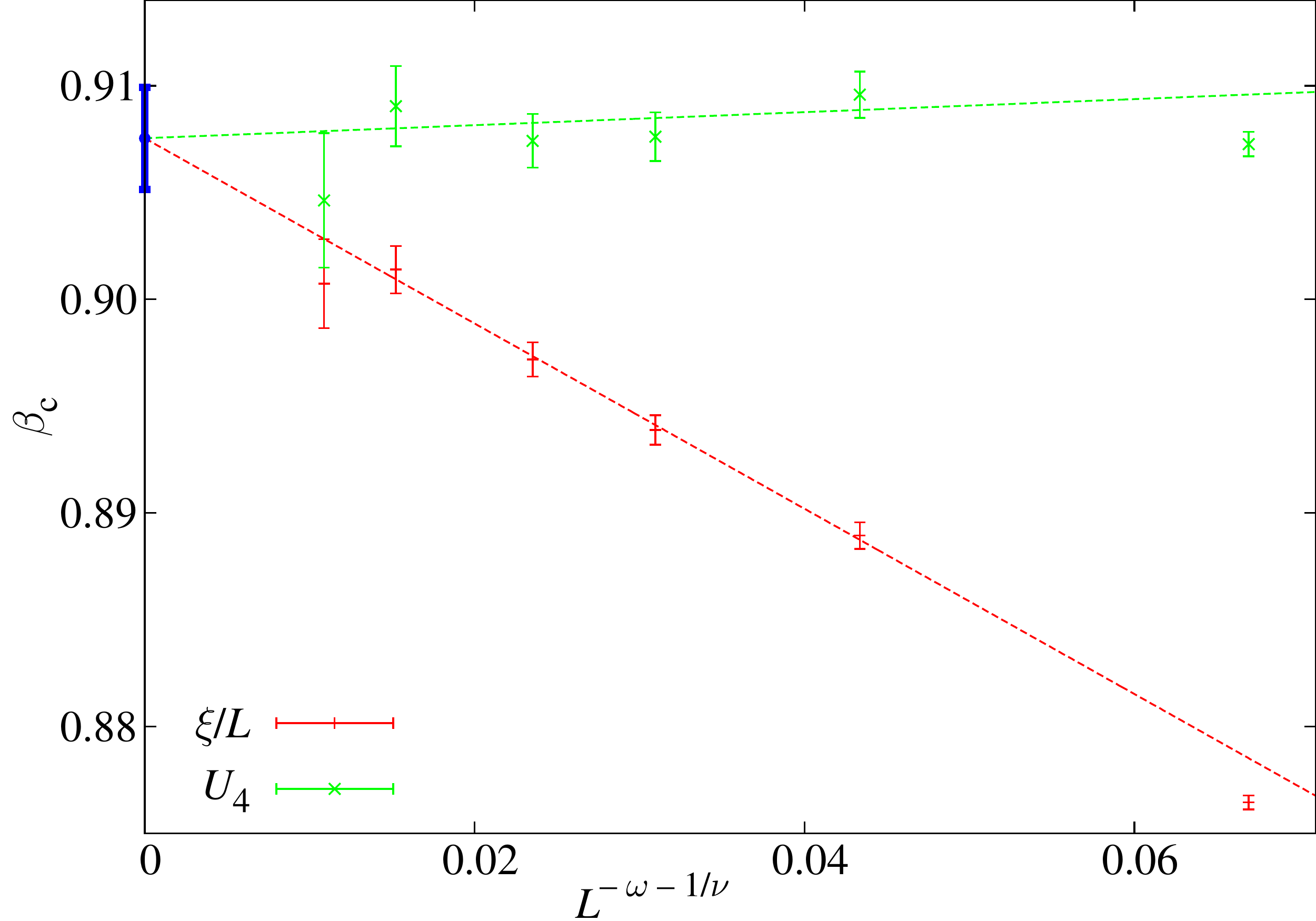}
\caption{(color online). Computation of $\beta_\text{c}$ with the quotients
  method.  We fit the crossing point $\beta^\text{cross}(L)$ of lattices
  $(L,2L)$ to $\beta^\text{cross} = \beta_\text{c} + A L^{-\omega-1/\nu}$,
  using both $U_4$ and $R_\xi=\xi/L$ to determine $\beta^\text{cross}$ and
  discarding the data for the $(6,12)$ crossing. The common extrapolated value
  is $\beta_\text{c} = 0.9075(11)[\overline{13}]$, where the first error bar
  is the statistical error in the fit while the second one is due to the error
  in $\omega+1/\nu$ (the overline denotes that $\beta_\text{c}$ is
  anticorrelated with $\omega+1/\nu$).\label{fig:betac}}
\end{figure}
As discussed in Section~\ref{sect:FSS}, the crossing point behaves as 
\begin{equation}\label{eq:Tc}
\beta^\text{cross}_g = \beta_\text{c} + A_{\beta_\text{c},g} L^{-\omega -
  1/\nu}+\ldots,
\end{equation}
where $g$ denotes the dimensionless quantity used to compute the crossing
points.\footnote{As usual, $\beta=1/T$. We employ it in order to allow for a
  direct comparison with raw data from Ref.~\onlinecite{hasenbusch:08b}.} We
can use this formula to determine the critical temperature of the system. To
this end, we perform a joint fit to Eq.~\eqref{eq:Tc} using the crossings computed
both with $U_4$ and with $R_\xi$ (where $\beta_\text{c}$ is a common fit
parameter). We take the value of $\omega+1/\nu$ from the fit in
Section~\ref{sect:mega-fit}. The final value, again fitting for $L\geq8$ is
\begin{align}
\beta_\text{c} &= 0.9075(11)[\overline{13}], & \chi^2/\text{d.o.f.} &= 6.15/7.
\end{align}
The first error bar is the statistical uncertainty in the fit and the second
error bar is due to our uncertainty in $\omega+1/\nu$.  The line over the
second error bar denotes that the estimate of $\beta_\text{c}$ is
anticorrelated with that of $\omega+1/\nu$.  The corresponding value for
$T_\text{c}$ is, therefore,
\begin{equation}\label{eq:Tcnum}
T_\text{c} = 1.1019(13)[16].
\end{equation}
The amplitudes $A_{\beta_\text{c}, g}$  are
\begin{align}
A_{\beta_\text{c}, \xi} &= -0.434(34)[\overline{58}], & A_{\beta_\text{c}, U_4} &= 0.031(37)[49].
\end{align}
\begin{table}[t]
\begin{ruledtabular}
\begin{tabular}{r@{\,=\,}lllc}
\multicolumn{2}{c}{Universal quantity}  & 
\multicolumn{1}{c}{$A_{g,\xi}$} & 
\multicolumn{1}{c}{$A_{g,U_4}$} &$\chi^2/\text{d.o.f.}$ \\
\hline
$U_{1111}^*$ & $0.47141(68)[\overline{70}]$ & $-0.0681(87)[\overline{61}]$ & $\hphantom{-}0.005(9)[10] $ & $7.72/7$  \\
$U_{22}^*$   & $0.76808(76)[\overline{84}]$ & $-0.085(10)[\overline{8}]$   & $-0.003(10)[10]$ & $8.32/7$  \\ 
$U_{111}^*$  & $0.44886(73)[\overline{77}]$ & $-0.0723(93)[\overline{62}]$ & $\hphantom{-}0.008(10)[11]$    & $7.86/7$  \\ 
$B_\chi^*$   & $2.4142(33)[\overline{18}]$   & $-0.044(42)[14]$  &
$\hphantom{-}0.36(4)[10]$      & $8.91/7$  
\end{tabular}
\end{ruledtabular}
\caption{Universal quantities at the critical point.  We remind the reader
  that the first error bar is the statistical error in each fit, while the
  second is the effect of the uncertainty in our estimate of $\omega$ (we add
  an overline if the quantity is anticorrelated with $\omega$).  We also give
  the (non-universal) amplitudes $A_{g,\xi}$ and $A_{g,U_4}$ in the
  fits.  \label{tab:cumulants}}
\end{table}

\subsection{Dimensionless universal quantities}\label{sect:dimensionless}

As explained in Section~\ref{sect:mega-fit},  we have not used
the non-standard dimensionless
ratios defined in Section~\ref{sect:observables} [Eqs.~\eqref{eq:U22-def}--\eqref{eq:Bchi-def}] to determine the
critical exponents of the system. However, since some of these quantities
have been found useful in the past\cite{billoire:11,janus:12} and since 
they are universal quantities further characterizing the Ising spin glass
universality class, we have found it interesting to report 
their critical values.

We perform fits to 
\begin{equation}\label{eq:g-L-fit}
g_L^\text{cross} = g^* + A_g L^{-\omega},
\end{equation}
where $g$ is each of $U_{1111}, U_{111}, U_{22}, R_{12}$ and $B_\chi$.  We
take $\omega$ from Eq.~\eqref{eq:critical-parameters}. In all cases we include
all data with $L\geq 8$. In order to improve our statistics, we consider for
each $g$ its scaling on the crossing point of both $U_4$ and $R_\xi$ (with
common extrapolation $g^*$). Table~\ref{tab:cumulants} displays the results
for all quantities but $R_{12}.$

In fact, we realized that $R_{12}$ deserves a special analysis when making the
consistency test alluded to in Sect.~\ref{sect:FSS}. We performed again the
fit in Eq.~\eqref{eq:g-L-fit}, but for $g_{2L}^\mathrm{cross}$ this time. If
subdominant scaling corrections are truly negligible, as we assume in
Eq.~\eqref{eq:g-L-fit}, the universal extrapolation $g^*$ must come out
compatible. The estimate of $g^*$ changed by less than one tenth of an error
bar for $U_{1111}, U_{111}$ and $U_{22}$. In the case of $B_{\chi}^*$ the
obtained result varied a full error bar (we obtained
$B_{\chi}^*=2.4218(35)[\overline{42}]$ in the fit with $g_{2L}$). Given the
data correlation, this difference might be significant, so we suggest
doubling the error for $B_{\chi}^*$ in Table~\ref{tab:cumulants} if one wants 
to be specially careful.

\begin{figure}[t]
\includegraphics[height=\linewidth, angle=270]{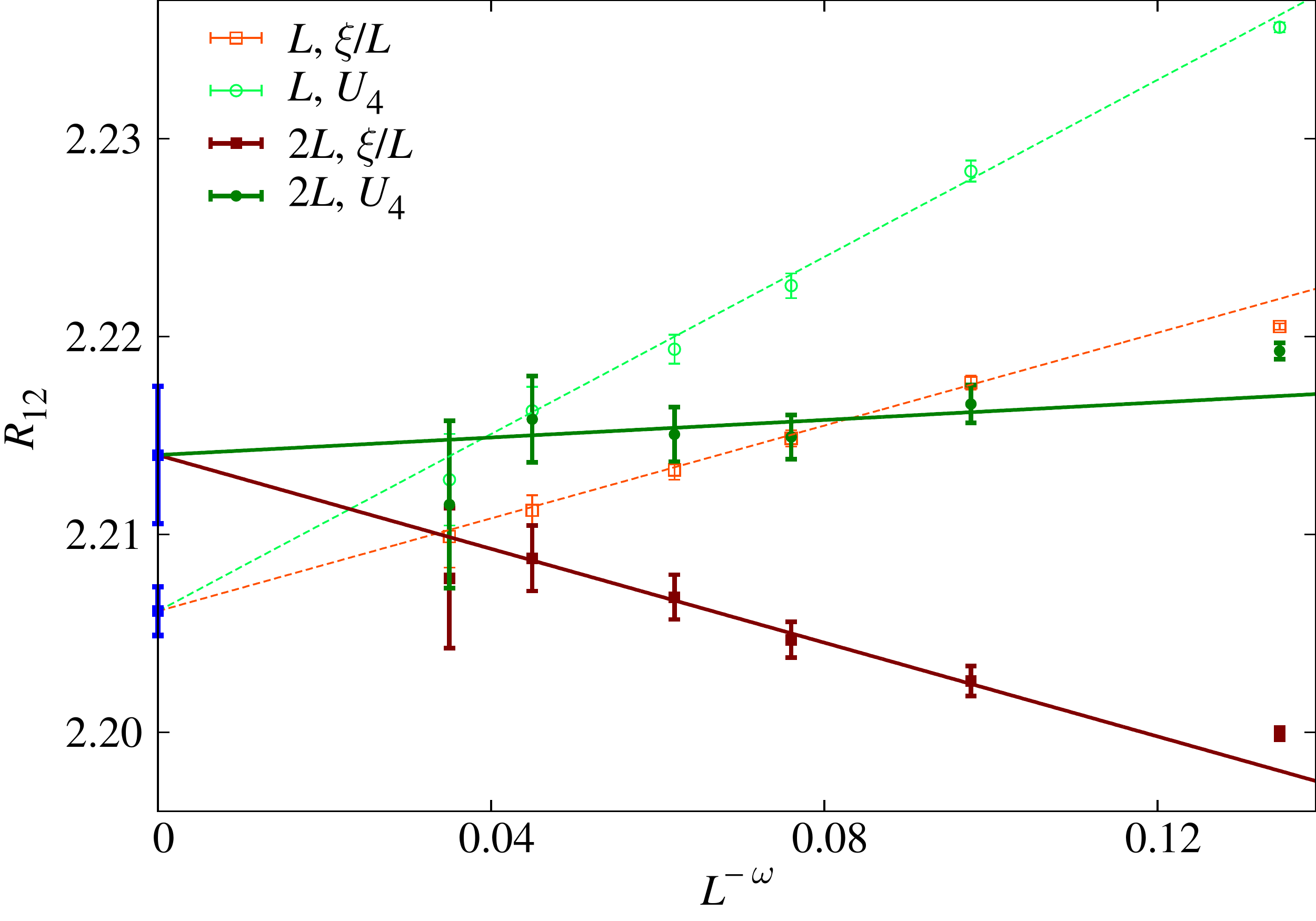}
\caption{(color online). Resolving the ambiguity in Eq.~\eqref{eq:g-L-fit}: at
  the crossing point of $\xi/L$ or $U_4$, one is free to consider the
  dimensionless quantity $g$ as computed for the small system
  ($g_L^\text{cross}$) or for the large system ($g_{2L}^\text{cross}$). This
  choice turns out to be immaterial for all quantities reported in
  Table~\ref{tab:cumulants}, but not for $R_{12}$. In the plot, we display the
  values of $R_{12}$ at the corresponding crossing point, as a function of
  $L^{-\omega}$. Empty (full) symbols correspond to the small (large) lattice
  in the pair $(L,2L)$ involved in the crossing. Lines are fits to
  Eq.~\eqref{eq:g-L-fit}, constrained to yield a common extrapolation for the
  $\xi/L$ crossings and for the $U_4$ crossings. The dashed (full) lines
  correspond to the fits for the small (large) lattices. The corresponding
  extrapolations are depicted in blue on the $L^{-\omega}=0$ axis. Both fits
  are performed for $L\geq 8$ and of good statistical quality:
  $\chi^2/\text{d.o.f.}=7.0/7$ (small lattice) and
  $\chi^2/\text{d.o.f.}=5.9/7$ (large lattice). In spite of this, both
  extrapolations are incompatible. Our final value for $R_{12}^*$,
  Eq.~\eqref{eq:R12}, corresponds to the minimal interval that includes both
  extrapolations and their statistical errors.
\label{fig:R12-ay-ay-ay}}
\end{figure}

Unfortunately, subleading scaling corrections are more difficult to control
for $R_{12}^*$. The extrapolation for $g_L$ and $g_{2L}$ are clearly
incompatible, see Figure~\ref{fig:R12-ay-ay-ay}. Considering subleading
corrections of order $L^{-2\omega}$ does not improve the situation. Therefore,
we have chosen a more conservative approach: we give as a final estimate the
interval covering both extrapolations and their errors
\begin{equation}\label{eq:R12}
R_{12}^*=2.211\pm 0.006\,.
\end{equation}
We emphasize that, when comparing with future work, it will be necessary to
keep in mind that the error in Eq.~\eqref{eq:R12} is of systematic rather than
of statistical nature.

\section{Conclusions}\label{sect:conclusions}

In this paper we have performed a finite-size scaling study of
the critical behavior of the Ising spin glass, using 
data from large-scale parallel tempering simulations 
performed on the Janus computer.  We have
followed a strategy based on the application of the quotients 
method and on the use of joint fits of several quantities
to obtain accurate estimates of all the critical exponents
of the system. We have also computed the critical value of 
several universal dimensionless quantities, as well as the 
value of the critical temperature. 

\section*{Acknowledgments}

The total simulation time devoted to this project was the equivalent
of 107 days of the full Janus machine ($L\geq 32$) and 43 days
of the full (3072 cores) \emph{Memento} cluster. For information about
both machines, see also \href{http://bifi.es}{http://bifi.es}.

The Janus project has been partially supported by the EU (FEDER funds, No.
UNZA05-33-003, MEC-DGA, Spain); by the European Research Council under the
European Union's Seventh Framework Programme (FP7/2007-2013, ERC grant
agreement no.  247328); by the MICINN (Spain) (contracts FIS2006-08533,
FIS2012-35719-C02, FIS2010-16587, TEC2010-19207); by the SUMA project of INFN
(Italy); by the Junta de Extremadura (GR10158); by the Microsoft Prize 2007
and by the European Union (PIRSES-GA-2011-295302).  F.R.-T. was supported by
the Italian Research Ministry through the FIRB project No. RBFR086NN1;
M.B.-J. was supported by the FPU program (Ministerio de Educacion, Spain);
R.A.B. and J.M.-G. were supported by the FPI program (Diputacion de Aragon,
Spain); finally J.M.G.-N. was
supported by the FPI program (Ministerio de Ciencia e Innovacion, Spain).

\begin{figure}[!t]
\includegraphics[height=\linewidth, angle=270]{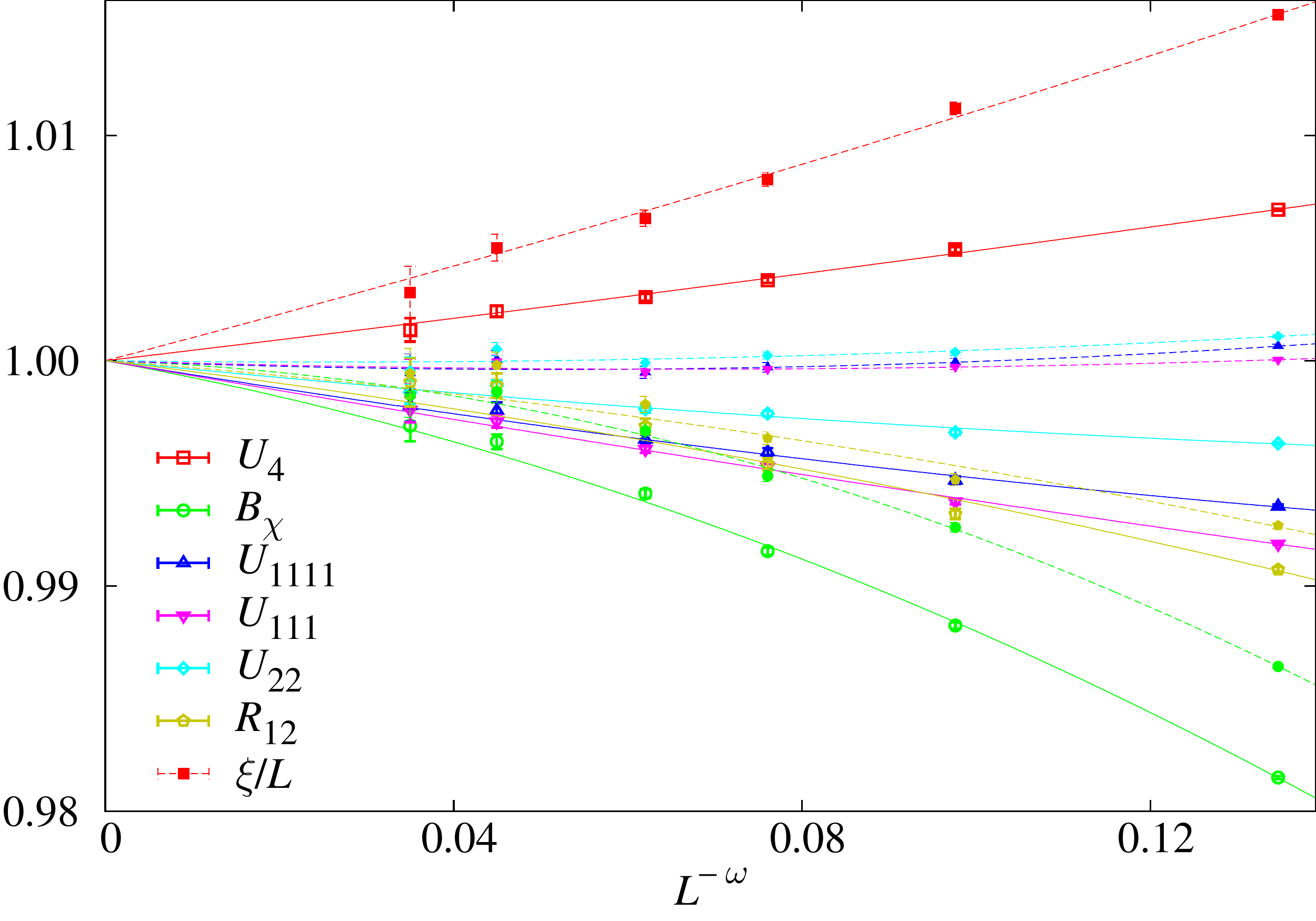}
\caption{(color online). Quotients of dimensionless quantities 
computed at the crossing points of $\xi/L$ (open symbols)
and of $U_4$ (filled symbols).  We also
plot individual fits to~\eqref{eq:dimensionless-quotients}
(solid lines for $\xi/L$ and dotted lines for $U_4$).
\label{fig:omega-alt}}
\end{figure}
\bigskip
\appendix

\section{Alternative computation of exponent {\boldmath $\omega$}}
\label{sec:omega-alt}

As mentioned in section~\ref{sect:mega-fit}, the computation 
of the scaling corrections exponent $\omega$ is the most 
delicate step in the analysis. Therefore, we present in this 
appendix an alternative way of approaching it, as a consistency
check of our results.

We start by making the rather obvious remark that the quotient $\mathcal{Q}_g=g_{2L}/g_L$
of any dimensionless quantity  $g$ at the crossing points $\beta^{\text{cross}}$
defined by any other dimensionless quantity $h$ [so $h_L(\beta^\text{cross}) =
h_{2L}(\beta_\text{cross})$] goes to one as the system size increases 
\begin{equation}\label{eq:dimensionless-quotients}
\mathcal Q^{\text{cross}}_g(L) = 1 + A_g L^{-\omega} + B_g L^{-2\omega} + \ldots,
\end{equation} 
The advantage of this equation is that, unlike  in our analysis of
Section~\ref{sect:mega-fit}, the asymptotic value is not another
parameter in the fit, but known in advance. We show $\mathcal Q^\text{cross}_g$
in Figure~\ref{fig:omega-alt} for $g=U_4, R_\xi, U_{1111}, U_{111}, U_{22}, B_\chi$
and $R_{12}$. In all cases we compute $\mathcal Q^{\text{cross}}$ at the crossing
points of both $U_4$ and $R_\xi$ (except for these two quantities, which
are obviously considered only at each other's crossing points).

In Figure~\ref{fig:omega-alt} we have performed individual fits
to~\eqref{eq:dimensionless-quotients} for each quantity, using the previously
computed value of $\omega$, to show that this description is consistent.  From
the plot we can see that some of the quantities (such as $R_{12}$) have very
clear subleading scaling corrections, but that for others the leading term
in~\eqref{eq:dimensionless-quotients} is quite sufficient.

Armed with this qualitative observation, we can do a second fit 
to~\eqref{eq:dimensionless-quotients}, this time leaving $\omega$
free and considering several dimensionless quantities at the same
time. In particular, in order to avoid the quadratic term, we have
considered $U_4, R_\xi, U_{1111}$ and $U_{111}$ and discarded the data for $L=6$.
 The result of this joint fit is 
\begin{align}
\omega &= 1.187(68), &   \chi^2/\text{d.o.f.} = 19.80/23,
\end{align}
compatible with our previous determination of $\omega$.


\begin{thebibliography}{51}%
\makeatletter
\providecommand \@ifxundefined [1]{%
 \@ifx{#1\undefined}
}%
\providecommand \@ifnum [1]{%
 \ifnum #1\expandafter \@firstoftwo
 \else \expandafter \@secondoftwo
 \fi
}%
\providecommand \@ifx [1]{%
 \ifx #1\expandafter \@firstoftwo
 \else \expandafter \@secondoftwo
 \fi
}%
\providecommand \natexlab [1]{#1}%
\providecommand \enquote  [1]{``#1''}%
\providecommand \bibnamefont  [1]{#1}%
\providecommand \bibfnamefont [1]{#1}%
\providecommand \citenamefont [1]{#1}%
\providecommand \href@noop [0]{\@secondoftwo}%
\providecommand \href [0]{\begingroup \@sanitize@url \@href}%
\providecommand \@href[1]{\@@startlink{#1}\@@href}%
\providecommand \@@href[1]{\endgroup#1\@@endlink}%
\providecommand \@sanitize@url [0]{\catcode `\\12\catcode `\$12\catcode
  `\&12\catcode `\#12\catcode `\^12\catcode `\_12\catcode `\%12\relax}%
\providecommand \@@startlink[1]{}%
\providecommand \@@endlink[0]{}%
\providecommand \url  [0]{\begingroup\@sanitize@url \@url }%
\providecommand \@url [1]{\endgroup\@href {#1}{\urlprefix }}%
\providecommand \urlprefix  [0]{URL }%
\providecommand \Eprint [0]{\href }%
\providecommand \doibase [0]{http://dx.doi.org/}%
\providecommand \selectlanguage [0]{\@gobble}%
\providecommand \bibinfo  [0]{\@secondoftwo}%
\providecommand \bibfield  [0]{\@secondoftwo}%
\providecommand \translation [1]{[#1]}%
\providecommand \BibitemOpen [0]{}%
\providecommand \bibitemStop [0]{}%
\providecommand \bibitemNoStop [0]{.\EOS\space}%
\providecommand \EOS [0]{\spacefactor3000\relax}%
\providecommand \BibitemShut  [1]{\csname bibitem#1\endcsname}%
\let\auto@bib@innerbib\@empty
\bibitem [{\citenamefont {Mydosh}(1993)}]{mydosh:93}%
  \BibitemOpen
  \bibfield  {author} {\bibinfo {author} {\bibfnamefont {J.~A.}\ \bibnamefont
  {Mydosh}},\ }\href@noop {} {\emph {\bibinfo {title} {Spin Glasses: an
  Experimental Introduction}}}\ (\bibinfo  {publisher} {Taylor and Francis},\
  \bibinfo {address} {London},\ \bibinfo {year} {1993})\BibitemShut {NoStop}%
\bibitem [{\citenamefont {M{\'e}zard}\ \emph {et~al.}(1987)\citenamefont
  {M{\'e}zard}, \citenamefont {Parisi},\ and\ \citenamefont
  {Virasoro}}]{mezard:87}%
  \BibitemOpen
  \bibfield  {author} {\bibinfo {author} {\bibfnamefont {M.}~\bibnamefont
  {M{\'e}zard}}, \bibinfo {author} {\bibfnamefont {G.}~\bibnamefont {Parisi}},
  \ and\ \bibinfo {author} {\bibfnamefont {M.}~\bibnamefont {Virasoro}},\
  }\href@noop {} {\emph {\bibinfo {title} {Spin-Glass Theory and Beyond}}}\
  (\bibinfo  {publisher} {World Scientific},\ \bibinfo {address} {Singapore},\
  \bibinfo {year} {1987})\BibitemShut {NoStop}%
\bibitem [{\citenamefont {Edwards}\ and\ \citenamefont
  {Anderson}(1975)}]{edwards:75}%
  \BibitemOpen
  \bibfield  {author} {\bibinfo {author} {\bibfnamefont {S.~F.}\ \bibnamefont
  {Edwards}}\ and\ \bibinfo {author} {\bibfnamefont {P.~W.}\ \bibnamefont
  {Anderson}},\ }\href {\doibase 10.1088/0305-4608/5/5/017} {\bibfield
  {journal} {\bibinfo  {journal} {J. Phys. F}\ }\textbf {\bibinfo {volume}
  {5}},\ \bibinfo {pages} {965} (\bibinfo {year} {1975})}\BibitemShut {NoStop}%
\bibitem [{\citenamefont {Palassini}\ and\ \citenamefont
  {Caracciolo}(1999)}]{palassini:99}%
  \BibitemOpen
  \bibfield  {author} {\bibinfo {author} {\bibfnamefont {M.}~\bibnamefont
  {Palassini}}\ and\ \bibinfo {author} {\bibfnamefont {S.}~\bibnamefont
  {Caracciolo}},\ }\href {\doibase 10.1103/PhysRevLett.82.5128} {\bibfield
  {journal} {\bibinfo  {journal} {Phys. Rev. Lett.}\ }\textbf {\bibinfo
  {volume} {82}},\ \bibinfo {pages} {5128} (\bibinfo {year} {1999})},\ \Eprint
  {http://arxiv.org/abs/arXiv:cond-mat/9904246} {arXiv:cond-mat/9904246}
  \BibitemShut {NoStop}%
\bibitem [{\citenamefont {Ballesteros}\ \emph {et~al.}(2000)\citenamefont
  {Ballesteros}, \citenamefont {Cruz}, \citenamefont {Fernandez}, \citenamefont
  {Martin-Mayor}, \citenamefont {Pech}, \citenamefont {Ruiz-Lorenzo},
  \citenamefont {Tarancon}, \citenamefont {Tellez}, \citenamefont {Ullod},\
  and\ \citenamefont {Ungil}}]{ballesteros:00}%
  \BibitemOpen
  \bibfield  {author} {\bibinfo {author} {\bibfnamefont {H.~G.}\ \bibnamefont
  {Ballesteros}}, \bibinfo {author} {\bibfnamefont {A.}~\bibnamefont {Cruz}},
  \bibinfo {author} {\bibfnamefont {L.~A.}\ \bibnamefont {Fernandez}}, \bibinfo
  {author} {\bibfnamefont {V.}~\bibnamefont {Martin-Mayor}}, \bibinfo {author}
  {\bibfnamefont {J.}~\bibnamefont {Pech}}, \bibinfo {author} {\bibfnamefont
  {J.~J.}\ \bibnamefont {Ruiz-Lorenzo}}, \bibinfo {author} {\bibfnamefont
  {A.}~\bibnamefont {Tarancon}}, \bibinfo {author} {\bibfnamefont
  {P.}~\bibnamefont {Tellez}}, \bibinfo {author} {\bibfnamefont {C.~L.}\
  \bibnamefont {Ullod}}, \ and\ \bibinfo {author} {\bibfnamefont
  {C.}~\bibnamefont {Ungil}},\ }\href {\doibase 10.1103/PhysRevB.62.14237}
  {\bibfield  {journal} {\bibinfo  {journal} {Phys. Rev. B}\ }\textbf {\bibinfo
  {volume} {62}},\ \bibinfo {pages} {14237} (\bibinfo {year} {2000})},\ \Eprint
  {http://arxiv.org/abs/arXiv:cond-mat/0006211} {arXiv:cond-mat/0006211}
  \BibitemShut {NoStop}%
\bibitem [{\citenamefont {Kawashima}\ and\ \citenamefont
  {Young}(1996)}]{kawashima:96}%
  \BibitemOpen
  \bibfield  {author} {\bibinfo {author} {\bibfnamefont {N.}~\bibnamefont
  {Kawashima}}\ and\ \bibinfo {author} {\bibfnamefont {A.~P.}\ \bibnamefont
  {Young}},\ }\href {\doibase 10.1103/PhysRevB.53.R484} {\bibfield  {journal}
  {\bibinfo  {journal} {Phys. Rev. B}\ }\textbf {\bibinfo {volume} {53}},\
  \bibinfo {pages} {R484} (\bibinfo {year} {1996})}\BibitemShut {NoStop}%
\bibitem [{\citenamefont {I{\~n}iguez}\ \emph {et~al.}(1996)\citenamefont
  {I{\~n}iguez}, \citenamefont {Parisi},\ and\ \citenamefont
  {Ruiz-Lorenzo}}]{iniguez:96}%
  \BibitemOpen
  \bibfield  {author} {\bibinfo {author} {\bibfnamefont {D.}~\bibnamefont
  {I{\~n}iguez}}, \bibinfo {author} {\bibfnamefont {G.}~\bibnamefont {Parisi}},
  \ and\ \bibinfo {author} {\bibfnamefont {J.~J.}\ \bibnamefont
  {Ruiz-Lorenzo}},\ }\href {\doibase 10.1088/0305-4470/29/15/009} {\bibfield
  {journal} {\bibinfo  {journal} {J. Phys. A: Math. and Gen.}\ }\textbf
  {\bibinfo {volume} {29}},\ \bibinfo {pages} {4337} (\bibinfo {year}
  {1996})}\BibitemShut {NoStop}%
\bibitem [{\citenamefont {Marinari}\ \emph {et~al.}(1998)\citenamefont
  {Marinari}, \citenamefont {Parisi},\ and\ \citenamefont
  {Ruiz-Lorenzo}}]{marinari:98d}%
  \BibitemOpen
  \bibfield  {author} {\bibinfo {author} {\bibfnamefont {E.}~\bibnamefont
  {Marinari}}, \bibinfo {author} {\bibfnamefont {G.}~\bibnamefont {Parisi}}, \
  and\ \bibinfo {author} {\bibfnamefont {J.~J.}\ \bibnamefont {Ruiz-Lorenzo}},\
  }\href {\doibase 10.1103/PhysRevB.58.14852} {\bibfield  {journal} {\bibinfo
  {journal} {Phys. Rev. B}\ }\textbf {\bibinfo {volume} {58}},\ \bibinfo
  {pages} {14852} (\bibinfo {year} {1998})}\BibitemShut {NoStop}%
\bibitem [{\citenamefont {I{\~n}iguez}\ \emph {et~al.}(1997)\citenamefont
  {I{\~n}iguez}, \citenamefont {Marinari}, \citenamefont {Parisi},\ and\
  \citenamefont {Ruiz-Lorenzo}}]{iniguez:97}%
  \BibitemOpen
  \bibfield  {author} {\bibinfo {author} {\bibfnamefont {D.}~\bibnamefont
  {I{\~n}iguez}}, \bibinfo {author} {\bibfnamefont {E.}~\bibnamefont
  {Marinari}}, \bibinfo {author} {\bibfnamefont {G.}~\bibnamefont {Parisi}}, \
  and\ \bibinfo {author} {\bibfnamefont {J.~J.}\ \bibnamefont {Ruiz-Lorenzo}},\
  }\href {\doibase 10.1088/0305-4470/30/21/010} {\bibfield  {journal} {\bibinfo
   {journal} {J. Phys. A: Math. and Gen.}\ }\textbf {\bibinfo {volume} {30}},\
  \bibinfo {pages} {7337} (\bibinfo {year} {1997})}\BibitemShut {NoStop}%
\bibitem [{\citenamefont {Berg}\ and\ \citenamefont {Janke}(1998)}]{berg:98}%
  \BibitemOpen
  \bibfield  {author} {\bibinfo {author} {\bibfnamefont {B.~A.}\ \bibnamefont
  {Berg}}\ and\ \bibinfo {author} {\bibfnamefont {W.}~\bibnamefont {Janke}},\
  }\href {\doibase 10.1103/PhysRevLett.80.4771} {\bibfield  {journal} {\bibinfo
   {journal} {Phys. Rev. Lett.}\ }\textbf {\bibinfo {volume} {80}},\ \bibinfo
  {pages} {4771} (\bibinfo {year} {1998})}\BibitemShut {NoStop}%
\bibitem [{\citenamefont {Janke}\ \emph {et~al.}(1998)\citenamefont {Janke},
  \citenamefont {Berg},\ and\ \citenamefont {Billoire}}]{janke:98b}%
  \BibitemOpen
  \bibfield  {author} {\bibinfo {author} {\bibfnamefont {W.}~\bibnamefont
  {Janke}}, \bibinfo {author} {\bibfnamefont {B.~A.}\ \bibnamefont {Berg}}, \
  and\ \bibinfo {author} {\bibfnamefont {A.}~\bibnamefont {Billoire}},\ }\href
  {\doibase 10.1002/(SICI)1521-3889(199811)7:5/6<544::AID-ANDP544>3.0.CO;2-1}
  {\bibfield  {journal} {\bibinfo  {journal} {Ann. Phys.}\ }\textbf {\bibinfo
  {volume} {7}},\ \bibinfo {pages} {544} (\bibinfo {year} {1998})}\BibitemShut
  {NoStop}%
\bibitem [{\citenamefont {Gunnarsson}\ \emph {et~al.}(1991)\citenamefont
  {Gunnarsson}, \citenamefont {Svendlindh}, \citenamefont {Nordblad},
  \citenamefont {Lundgren}, \citenamefont {Aruga},\ and\ \citenamefont
  {Ito}}]{gunnarsson:91}%
  \BibitemOpen
  \bibfield  {author} {\bibinfo {author} {\bibfnamefont {K.}~\bibnamefont
  {Gunnarsson}}, \bibinfo {author} {\bibfnamefont {P.}~\bibnamefont
  {Svendlindh}}, \bibinfo {author} {\bibfnamefont {P.}~\bibnamefont
  {Nordblad}}, \bibinfo {author} {\bibfnamefont {L.}~\bibnamefont {Lundgren}},
  \bibinfo {author} {\bibfnamefont {H.}~\bibnamefont {Aruga}}, \ and\ \bibinfo
  {author} {\bibfnamefont {A.}~\bibnamefont {Ito}},\ }\href {\doibase
  10.1103/PhysRevB.43.8199} {\bibfield  {journal} {\bibinfo  {journal} {Phys.
  Rev. B}\ }\textbf {\bibinfo {volume} {43}},\ \bibinfo {pages} {8199}
  (\bibinfo {year} {1991})}\BibitemShut {NoStop}%
\bibitem [{\citenamefont {Mari}\ and\ \citenamefont
  {Campbell}(2002)}]{mari:02}%
  \BibitemOpen
  \bibfield  {author} {\bibinfo {author} {\bibfnamefont {P.~O.}\ \bibnamefont
  {Mari}}\ and\ \bibinfo {author} {\bibfnamefont {I.~A.}\ \bibnamefont
  {Campbell}},\ }\href {\doibase 10.1103/PhysRevB.65.184409} {\bibfield
  {journal} {\bibinfo  {journal} {Phys. Rev. B}\ }\textbf {\bibinfo {volume}
  {65}},\ \bibinfo {pages} {184409} (\bibinfo {year} {2002})}\BibitemShut
  {NoStop}%
\bibitem [{\citenamefont {Nakamura}\ \emph {et~al.}(2003)\citenamefont
  {Nakamura}, \citenamefont {Endoh},\ and\ \citenamefont
  {Yamamoto}}]{nakamura:03}%
  \BibitemOpen
  \bibfield  {author} {\bibinfo {author} {\bibfnamefont {T.}~\bibnamefont
  {Nakamura}}, \bibinfo {author} {\bibfnamefont {S.-i.}\ \bibnamefont {Endoh}},
  \ and\ \bibinfo {author} {\bibfnamefont {T.}~\bibnamefont {Yamamoto}},\
  }\href {\doibase 10.1088/0305-4470/36/43/015} {\bibfield  {journal} {\bibinfo
   {journal} {J. Phys. A}\ }\textbf {\bibinfo {volume} {36}},\ \bibinfo {pages}
  {10895} (\bibinfo {year} {2003})}\BibitemShut {NoStop}%
\bibitem [{\citenamefont {Daboul}\ \emph {et~al.}(2004)\citenamefont {Daboul},
  \citenamefont {Chang},\ and\ \citenamefont {Aharony}}]{daboul:04}%
  \BibitemOpen
  \bibfield  {author} {\bibinfo {author} {\bibfnamefont {D.}~\bibnamefont
  {Daboul}}, \bibinfo {author} {\bibfnamefont {I.}~\bibnamefont {Chang}}, \
  and\ \bibinfo {author} {\bibfnamefont {A.}~\bibnamefont {Aharony}},\ }\href
  {\doibase 10.1140/epjb/e2004-00315-6} {\bibfield  {journal} {\bibinfo
  {journal} {Eur. Phys. J. B}\ }\textbf {\bibinfo {volume} {41}},\ \bibinfo
  {pages} {231} (\bibinfo {year} {2004})}\BibitemShut {NoStop}%
\bibitem [{\citenamefont {Pleimling}\ and\ \citenamefont
  {Campbell}(2005)}]{pleimling:05}%
  \BibitemOpen
  \bibfield  {author} {\bibinfo {author} {\bibfnamefont {M.}~\bibnamefont
  {Pleimling}}\ and\ \bibinfo {author} {\bibfnamefont {I.~A.}\ \bibnamefont
  {Campbell}},\ }\href {\doibase 10.1103/PhysRevB.72.184429} {\bibfield
  {journal} {\bibinfo  {journal} {Phys. Rev. B}\ }\textbf {\bibinfo {volume}
  {72}},\ \bibinfo {pages} {184429} (\bibinfo {year} {2005})}\BibitemShut
  {NoStop}%
\bibitem [{\citenamefont {Perez-Gaviro}\ \emph {et~al.}(2006)\citenamefont
  {Perez-Gaviro}, \citenamefont {Ruiz-Lorenzo},\ and\ \citenamefont
  {Taranc{\'o}n}}]{perez-gaviro:06}%
  \BibitemOpen
  \bibfield  {author} {\bibinfo {author} {\bibfnamefont {S.}~\bibnamefont
  {Perez-Gaviro}}, \bibinfo {author} {\bibfnamefont {J.~J.}\ \bibnamefont
  {Ruiz-Lorenzo}}, \ and\ \bibinfo {author} {\bibfnamefont {A.}~\bibnamefont
  {Taranc{\'o}n}},\ }\href {\doibase 10.1088/0305-4470/39/27/001} {\bibfield
  {journal} {\bibinfo  {journal} {J. Phys. A: Math. Gen.}\ }\textbf {\bibinfo
  {volume} {39}},\ \bibinfo {pages} {8567} (\bibinfo {year}
  {2006})}\BibitemShut {NoStop}%
\bibitem [{\citenamefont {Parisen~Toldin}\ \emph {et~al.}(2006)\citenamefont
  {Parisen~Toldin}, \citenamefont {Pelissetto},\ and\ \citenamefont
  {Vicari}}]{parisen:06}%
  \BibitemOpen
  \bibfield  {author} {\bibinfo {author} {\bibfnamefont {F.}~\bibnamefont
  {Parisen~Toldin}}, \bibinfo {author} {\bibfnamefont {A.}~\bibnamefont
  {Pelissetto}}, \ and\ \bibinfo {author} {\bibfnamefont {E.}~\bibnamefont
  {Vicari}},\ }\href {\doibase 10.1088/1742-5468/2006/06/P06002} {\bibfield
  {journal} {\bibinfo  {journal} {J. Stat. Mech.: Theory Exp.}\ ,\ \bibinfo
  {pages} {P06002}} (\bibinfo {year} {2006})}\BibitemShut {NoStop}%
\bibitem [{\citenamefont {J{\"o}rg}(2006)}]{jorg:06}%
  \BibitemOpen
  \bibfield  {author} {\bibinfo {author} {\bibfnamefont {T.}~\bibnamefont
  {J{\"o}rg}},\ }\href {\doibase 10.1103/PhysRevB.73.224431} {\bibfield
  {journal} {\bibinfo  {journal} {Phys. Rev. B}\ }\textbf {\bibinfo {volume}
  {73}},\ \bibinfo {pages} {224431} (\bibinfo {year} {2006})}\BibitemShut
  {NoStop}%
\bibitem [{\citenamefont {Campbell}\ \emph {et~al.}(2006)\citenamefont
  {Campbell}, \citenamefont {Hukushima},\ and\ \citenamefont
  {Takayama}}]{campbell:06}%
  \BibitemOpen
  \bibfield  {author} {\bibinfo {author} {\bibfnamefont {I.~A.}\ \bibnamefont
  {Campbell}}, \bibinfo {author} {\bibfnamefont {K.}~\bibnamefont {Hukushima}},
  \ and\ \bibinfo {author} {\bibfnamefont {H.}~\bibnamefont {Takayama}},\
  }\href {\doibase 10.1103/PhysRevLett.97.117202} {\bibfield  {journal}
  {\bibinfo  {journal} {Phys. Rev. Lett.}\ }\textbf {\bibinfo {volume} {97}},\
  \bibinfo {pages} {117202} (\bibinfo {year} {2006})}\BibitemShut {NoStop}%
\bibitem [{\citenamefont {Katzgraber}\ \emph {et~al.}(2006)\citenamefont
  {Katzgraber}, \citenamefont {K\"orner},\ and\ \citenamefont
  {Young}}]{katzgraber:06}%
  \BibitemOpen
  \bibfield  {author} {\bibinfo {author} {\bibfnamefont {H.~G.}\ \bibnamefont
  {Katzgraber}}, \bibinfo {author} {\bibfnamefont {M.}~\bibnamefont
  {K\"orner}}, \ and\ \bibinfo {author} {\bibfnamefont {A.~P.}\ \bibnamefont
  {Young}},\ }\href {\doibase 10.1103/PhysRevB.73.224432} {\bibfield  {journal}
  {\bibinfo  {journal} {Phys. Rev. B}\ }\textbf {\bibinfo {volume} {73}},\
  \bibinfo {pages} {224432} (\bibinfo {year} {2006})}\BibitemShut {NoStop}%
\bibitem [{\citenamefont {Machta}\ \emph {et~al.}(2008)\citenamefont {Machta},
  \citenamefont {Newman},\ and\ \citenamefont {Stein}}]{machta:08}%
  \BibitemOpen
  \bibfield  {author} {\bibinfo {author} {\bibfnamefont {J.}~\bibnamefont
  {Machta}}, \bibinfo {author} {\bibfnamefont {C.~M.}\ \bibnamefont {Newman}},
  \ and\ \bibinfo {author} {\bibfnamefont {D.~L.}\ \bibnamefont {Stein}},\
  }\href@noop {} {\bibfield  {journal} {\bibinfo  {journal} {J. Stat. Phys.}\
  }\textbf {\bibinfo {volume} {130}},\ \bibinfo {pages} {113} (\bibinfo {year}
  {2008})}\BibitemShut {NoStop}%
\bibitem [{\citenamefont {Hasenbusch}\ \emph
  {et~al.}(2008{\natexlab{a}})\citenamefont {Hasenbusch}, \citenamefont
  {Pelissetto},\ and\ \citenamefont {Vicari}}]{hasenbusch:08}%
  \BibitemOpen
  \bibfield  {author} {\bibinfo {author} {\bibfnamefont {M.}~\bibnamefont
  {Hasenbusch}}, \bibinfo {author} {\bibfnamefont {A.}~\bibnamefont
  {Pelissetto}}, \ and\ \bibinfo {author} {\bibfnamefont {E.}~\bibnamefont
  {Vicari}},\ }\href {\doibase 10.1088/1742-5468/2008/02/L02001} {\bibfield
  {journal} {\bibinfo  {journal} {J. Stat. Mech.}\ ,\ \bibinfo {pages}
  {L02001}} (\bibinfo {year} {2008}{\natexlab{a}})}\BibitemShut {NoStop}%
\bibitem [{\citenamefont {Hasenbusch}\ \emph
  {et~al.}(2008{\natexlab{b}})\citenamefont {Hasenbusch}, \citenamefont
  {Pelissetto},\ and\ \citenamefont {Vicari}}]{hasenbusch:08b}%
  \BibitemOpen
  \bibfield  {author} {\bibinfo {author} {\bibfnamefont {M.}~\bibnamefont
  {Hasenbusch}}, \bibinfo {author} {\bibfnamefont {A.}~\bibnamefont
  {Pelissetto}}, \ and\ \bibinfo {author} {\bibfnamefont {E.}~\bibnamefont
  {Vicari}},\ }\href {\doibase 10.1103/PhysRevB.78.214205} {\bibfield
  {journal} {\bibinfo  {journal} {Phys. Rev. B}\ }\textbf {\bibinfo {volume}
  {78}},\ \bibinfo {pages} {214205} (\bibinfo {year}
  {2008}{\natexlab{b}})}\BibitemShut {NoStop}%
\bibitem [{\citenamefont {J{\"o}rg}\ and\ \citenamefont
  {Katzgraber}(2008)}]{jorg:08c}%
  \BibitemOpen
  \bibfield  {author} {\bibinfo {author} {\bibfnamefont {T.}~\bibnamefont
  {J{\"o}rg}}\ and\ \bibinfo {author} {\bibfnamefont {H.~G.}\ \bibnamefont
  {Katzgraber}},\ }\href {\doibase 10.1103/PhysRevB.77.214426} {\bibfield
  {journal} {\bibinfo  {journal} {Phys. Rev. B}\ }\textbf {\bibinfo {volume}
  {77}},\ \bibinfo {pages} {214426} (\bibinfo {year} {2008})},\ \Eprint
  {http://arxiv.org/abs/arXiv:0803.3339} {arXiv:0803.3339} \BibitemShut
  {NoStop}%
\bibitem [{\citenamefont {Belletti}\ \emph {et~al.}(2008)\citenamefont
  {Belletti}, \citenamefont {Cotallo}, \citenamefont {Cruz}, \citenamefont
  {Fernández}, \citenamefont {Gordillo}, \citenamefont {Maiorano},
  \citenamefont {Mantovani}, \citenamefont {Marinari}, \citenamefont
  {Martín-Mayor}, \citenamefont {Mu{\~n}oz~Sudupe}, \citenamefont {Navarro},
  \citenamefont {Pérez-Gaviro}, \citenamefont {Ruiz-Lorenzo}, \citenamefont
  {Schifano}, \citenamefont {Sciretti}, \citenamefont {Tarancón},
  \citenamefont {Tripiccione},\ and\ \citenamefont {Velasco}}]{janus:08}%
  \BibitemOpen
  \bibfield  {author} {\bibinfo {author} {\bibfnamefont {F.}~\bibnamefont
  {Belletti}}, \bibinfo {author} {\bibfnamefont {M.}~\bibnamefont {Cotallo}},
  \bibinfo {author} {\bibfnamefont {A.}~\bibnamefont {Cruz}}, \bibinfo {author}
  {\bibfnamefont {L.~A.}\ \bibnamefont {Fernández}}, \bibinfo {author}
  {\bibfnamefont {A.}~\bibnamefont {Gordillo}}, \bibinfo {author}
  {\bibfnamefont {A.}~\bibnamefont {Maiorano}}, \bibinfo {author}
  {\bibfnamefont {F.}~\bibnamefont {Mantovani}}, \bibinfo {author}
  {\bibfnamefont {E.}~\bibnamefont {Marinari}}, \bibinfo {author}
  {\bibfnamefont {V.}~\bibnamefont {Martín-Mayor}}, \bibinfo {author}
  {\bibfnamefont {A.}~\bibnamefont {Mu{\~n}oz~Sudupe}}, \bibinfo {author}
  {\bibfnamefont {D.}~\bibnamefont {Navarro}}, \bibinfo {author} {\bibfnamefont
  {S.}~\bibnamefont {Pérez-Gaviro}}, \bibinfo {author} {\bibfnamefont {J.~J.}\
  \bibnamefont {Ruiz-Lorenzo}}, \bibinfo {author} {\bibfnamefont {S.~F.}\
  \bibnamefont {Schifano}}, \bibinfo {author} {\bibfnamefont {D.}~\bibnamefont
  {Sciretti}}, \bibinfo {author} {\bibfnamefont {A.}~\bibnamefont {Tarancón}},
  \bibinfo {author} {\bibfnamefont {R.}~\bibnamefont {Tripiccione}}, \ and\
  \bibinfo {author} {\bibfnamefont {J.~L.}\ \bibnamefont {Velasco}} (\bibinfo
  {collaboration} {Janus Collaboration}),\ }\href {\doibase
  10.1016/j.cpc.2007.09.006} {\bibfield  {journal} {\bibinfo  {journal} {Comp.
  Phys. Comm.}\ }\textbf {\bibinfo {volume} {178}},\ \bibinfo {pages} {208}
  (\bibinfo {year} {2008})},\ \Eprint {http://arxiv.org/abs/arXiv:0704.3573}
  {arXiv:0704.3573} \BibitemShut {NoStop}%
\bibitem [{\citenamefont {Belletti}\ \emph {et~al.}(2009)\citenamefont
  {Belletti}, \citenamefont {Guidetti}, \citenamefont {Maiorano}, \citenamefont
  {Mantovani}, \citenamefont {Schifano}, \citenamefont {Tripiccione},
  \citenamefont {Cotallo}, \citenamefont {Perez-Gaviro}, \citenamefont
  {Sciretti}, \citenamefont {Velasco}, \citenamefont {Cruz}, \citenamefont
  {Navarro}, \citenamefont {Tarancon}, \citenamefont {Fernandez}, \citenamefont
  {Martin-Mayor}, \citenamefont {Mu{\~n}oz-Sudupe}, \citenamefont {Yllanes},
  \citenamefont {Gordillo-Guerrero}, \citenamefont {Ruiz-Lorenzo},
  \citenamefont {Marinari}, \citenamefont {Parisi}, \citenamefont {Rossi},\
  and\ \citenamefont {Zanier}}]{janus:09}%
  \BibitemOpen
  \bibfield  {author} {\bibinfo {author} {\bibfnamefont {F.}~\bibnamefont
  {Belletti}}, \bibinfo {author} {\bibfnamefont {M.}~\bibnamefont {Guidetti}},
  \bibinfo {author} {\bibfnamefont {A.}~\bibnamefont {Maiorano}}, \bibinfo
  {author} {\bibfnamefont {F.}~\bibnamefont {Mantovani}}, \bibinfo {author}
  {\bibfnamefont {S.~F.}\ \bibnamefont {Schifano}}, \bibinfo {author}
  {\bibfnamefont {R.}~\bibnamefont {Tripiccione}}, \bibinfo {author}
  {\bibfnamefont {M.}~\bibnamefont {Cotallo}}, \bibinfo {author} {\bibfnamefont
  {S.}~\bibnamefont {Perez-Gaviro}}, \bibinfo {author} {\bibfnamefont
  {D.}~\bibnamefont {Sciretti}}, \bibinfo {author} {\bibfnamefont {J.~L.}\
  \bibnamefont {Velasco}}, \bibinfo {author} {\bibfnamefont {A.}~\bibnamefont
  {Cruz}}, \bibinfo {author} {\bibfnamefont {D.}~\bibnamefont {Navarro}},
  \bibinfo {author} {\bibfnamefont {A.}~\bibnamefont {Tarancon}}, \bibinfo
  {author} {\bibfnamefont {L.~A.}\ \bibnamefont {Fernandez}}, \bibinfo {author}
  {\bibfnamefont {V.}~\bibnamefont {Martin-Mayor}}, \bibinfo {author}
  {\bibfnamefont {A.}~\bibnamefont {Mu{\~n}oz-Sudupe}}, \bibinfo {author}
  {\bibfnamefont {D.}~\bibnamefont {Yllanes}}, \bibinfo {author} {\bibfnamefont
  {A.}~\bibnamefont {Gordillo-Guerrero}}, \bibinfo {author} {\bibfnamefont
  {J.~J.}\ \bibnamefont {Ruiz-Lorenzo}}, \bibinfo {author} {\bibfnamefont
  {E.}~\bibnamefont {Marinari}}, \bibinfo {author} {\bibfnamefont
  {G.}~\bibnamefont {Parisi}}, \bibinfo {author} {\bibfnamefont
  {M.}~\bibnamefont {Rossi}}, \ and\ \bibinfo {author} {\bibfnamefont
  {G.}~\bibnamefont {Zanier}} (\bibinfo {collaboration} {Janus
  Collaboration}),\ }\href {\doibase 10.1109/MCSE.2009.11} {\bibfield
  {journal} {\bibinfo  {journal} {Computing in Science and Engineering}\
  }\textbf {\bibinfo {volume} {11}},\ \bibinfo {pages} {48} (\bibinfo {year}
  {2009})}\BibitemShut {NoStop}%
\bibitem [{\citenamefont {Alvarez~Ba{\~n}os}\ \emph
  {et~al.}(2010{\natexlab{a}})\citenamefont {Alvarez~Ba{\~n}os}, \citenamefont
  {Cruz}, \citenamefont {Fernandez}, \citenamefont {Gil-Narvion}, \citenamefont
  {Gordillo-Guerrero}, \citenamefont {Guidetti}, \citenamefont {Maiorano},
  \citenamefont {Mantovani}, \citenamefont {Marinari}, \citenamefont
  {Martin-Mayor}, \citenamefont {Monforte-Garcia}, \citenamefont
  {Mu{\~n}oz~Sudupe}, \citenamefont {Navarro}, \citenamefont {Parisi},
  \citenamefont {Perez-Gaviro}, \citenamefont {Ruiz-Lorenzo}, \citenamefont
  {Schifano}, \citenamefont {Seoane}, \citenamefont {Tarancon}, \citenamefont
  {Tripiccione},\ and\ \citenamefont {Yllanes}}]{janus:10b}%
  \BibitemOpen
  \bibfield  {author} {\bibinfo {author} {\bibfnamefont {R.}~\bibnamefont
  {Alvarez~Ba{\~n}os}}, \bibinfo {author} {\bibfnamefont {A.}~\bibnamefont
  {Cruz}}, \bibinfo {author} {\bibfnamefont {L.~A.}\ \bibnamefont {Fernandez}},
  \bibinfo {author} {\bibfnamefont {J.~M.}\ \bibnamefont {Gil-Narvion}},
  \bibinfo {author} {\bibfnamefont {A.}~\bibnamefont {Gordillo-Guerrero}},
  \bibinfo {author} {\bibfnamefont {M.}~\bibnamefont {Guidetti}}, \bibinfo
  {author} {\bibfnamefont {A.}~\bibnamefont {Maiorano}}, \bibinfo {author}
  {\bibfnamefont {F.}~\bibnamefont {Mantovani}}, \bibinfo {author}
  {\bibfnamefont {E.}~\bibnamefont {Marinari}}, \bibinfo {author}
  {\bibfnamefont {V.}~\bibnamefont {Martin-Mayor}}, \bibinfo {author}
  {\bibfnamefont {J.}~\bibnamefont {Monforte-Garcia}}, \bibinfo {author}
  {\bibfnamefont {A.}~\bibnamefont {Mu{\~n}oz~Sudupe}}, \bibinfo {author}
  {\bibfnamefont {D.}~\bibnamefont {Navarro}}, \bibinfo {author} {\bibfnamefont
  {G.}~\bibnamefont {Parisi}}, \bibinfo {author} {\bibfnamefont
  {S.}~\bibnamefont {Perez-Gaviro}}, \bibinfo {author} {\bibfnamefont {J.~J.}\
  \bibnamefont {Ruiz-Lorenzo}}, \bibinfo {author} {\bibfnamefont {S.~F.}\
  \bibnamefont {Schifano}}, \bibinfo {author} {\bibfnamefont {B.}~\bibnamefont
  {Seoane}}, \bibinfo {author} {\bibfnamefont {A.}~\bibnamefont {Tarancon}},
  \bibinfo {author} {\bibfnamefont {R.}~\bibnamefont {Tripiccione}}, \ and\
  \bibinfo {author} {\bibfnamefont {D.}~\bibnamefont {Yllanes}} (\bibinfo
  {collaboration} {Janus Collaboration}),\ }\href {\doibase
  10.1103/PhysRevLett.105.177202} {\bibfield  {journal} {\bibinfo  {journal}
  {Phys. Rev. Lett.}\ }\textbf {\bibinfo {volume} {105}},\ \bibinfo {pages}
  {177202} (\bibinfo {year} {2010}{\natexlab{a}})},\ \Eprint
  {http://arxiv.org/abs/arXiv:1003.2943} {arXiv:1003.2943} \BibitemShut
  {NoStop}%
\bibitem [{\citenamefont {Ba\~{n}os}\ \emph
  {et~al.}(2012{\natexlab{a}})\citenamefont {Ba\~{n}os}, \citenamefont {Cruz},
  \citenamefont {Fernandez}, \citenamefont {Gil-Narvion}, \citenamefont
  {Gordillo-Guerrero}, \citenamefont {Guidetti}, \citenamefont {Iniguez},
  \citenamefont {Maiorano}, \citenamefont {Marinari}, \citenamefont
  {Martin-Mayor}, \citenamefont {Monforte-Garcia}, \citenamefont
  {Muñoz~Sudupe}, \citenamefont {Navarro}, \citenamefont {Parisi},
  \citenamefont {Perez-Gaviro}, \citenamefont {Ruiz-Lorenzo}, \citenamefont
  {Schifano}, \citenamefont {Seoane}, \citenamefont {Tarancon}, \citenamefont
  {Tellez}, \citenamefont {Tripiccione},\ and\ \citenamefont
  {Yllanes}}]{janus:12}%
  \BibitemOpen
  \bibfield  {author} {\bibinfo {author} {\bibfnamefont {R.~A.}\ \bibnamefont
  {Ba\~{n}os}}, \bibinfo {author} {\bibfnamefont {A.}~\bibnamefont {Cruz}},
  \bibinfo {author} {\bibfnamefont {L.~A.}\ \bibnamefont {Fernandez}}, \bibinfo
  {author} {\bibfnamefont {J.~M.}\ \bibnamefont {Gil-Narvion}}, \bibinfo
  {author} {\bibfnamefont {A.}~\bibnamefont {Gordillo-Guerrero}}, \bibinfo
  {author} {\bibfnamefont {M.}~\bibnamefont {Guidetti}}, \bibinfo {author}
  {\bibfnamefont {D.}~\bibnamefont {Iniguez}}, \bibinfo {author} {\bibfnamefont
  {A.}~\bibnamefont {Maiorano}}, \bibinfo {author} {\bibfnamefont
  {E.}~\bibnamefont {Marinari}}, \bibinfo {author} {\bibfnamefont
  {V.}~\bibnamefont {Martin-Mayor}}, \bibinfo {author} {\bibfnamefont
  {J.}~\bibnamefont {Monforte-Garcia}}, \bibinfo {author} {\bibfnamefont
  {A.}~\bibnamefont {Muñoz~Sudupe}}, \bibinfo {author} {\bibfnamefont
  {D.}~\bibnamefont {Navarro}}, \bibinfo {author} {\bibfnamefont
  {G.}~\bibnamefont {Parisi}}, \bibinfo {author} {\bibfnamefont
  {S.}~\bibnamefont {Perez-Gaviro}}, \bibinfo {author} {\bibfnamefont {J.~J.}\
  \bibnamefont {Ruiz-Lorenzo}}, \bibinfo {author} {\bibfnamefont {S.~F.}\
  \bibnamefont {Schifano}}, \bibinfo {author} {\bibfnamefont {B.}~\bibnamefont
  {Seoane}}, \bibinfo {author} {\bibfnamefont {A.}~\bibnamefont {Tarancon}},
  \bibinfo {author} {\bibfnamefont {P.}~\bibnamefont {Tellez}}, \bibinfo
  {author} {\bibfnamefont {R.}~\bibnamefont {Tripiccione}}, \ and\ \bibinfo
  {author} {\bibfnamefont {D.}~\bibnamefont {Yllanes}},\ }\href {\doibase
  10.1073/pnas.1203295109} {\bibfield  {journal} {\bibinfo  {journal} {Proc.
  Natl. Acad. Sci. USA}\ }\textbf {\bibinfo {volume} {{109}}},\ \bibinfo
  {pages} {6452} (\bibinfo {year} {{2012}}{\natexlab{a}})}\BibitemShut
  {NoStop}%
\bibitem [{\citenamefont {Baity-Jesi}\ \emph
  {et~al.}(2013{\natexlab{a}})\citenamefont {Baity-Jesi}, \citenamefont
  {Ba\~{n}os}, \citenamefont {Cruz}, \citenamefont {Fernandez}, \citenamefont
  {Gil-Narvion}, \citenamefont {Gordillo-Guerrero}, \citenamefont {Iniguez},
  \citenamefont {Maiorano}, \citenamefont {F.}, \citenamefont {Marinari},
  \citenamefont {Martin-Mayor}, \citenamefont {Monforte-Garcia}, \citenamefont
  {Muñoz~Sudupe}, \citenamefont {Navarro}, \citenamefont {Parisi},
  \citenamefont {Perez-Gaviro}, \citenamefont {Pivanti}, \citenamefont
  {Ricci-Tersenghi}, \citenamefont {Ruiz-Lorenzo}, \citenamefont {Schifano},
  \citenamefont {Seoane}, \citenamefont {Tarancon}, \citenamefont
  {Tripiccione},\ and\ \citenamefont {Yllanes}}]{janus:13}%
  \BibitemOpen
  \bibfield  {author} {\bibinfo {author} {\bibfnamefont {M.}~\bibnamefont
  {Baity-Jesi}}, \bibinfo {author} {\bibfnamefont {R.~A.}\ \bibnamefont
  {Ba\~{n}os}}, \bibinfo {author} {\bibfnamefont {A.}~\bibnamefont {Cruz}},
  \bibinfo {author} {\bibfnamefont {L.~A.}\ \bibnamefont {Fernandez}}, \bibinfo
  {author} {\bibfnamefont {J.~M.}\ \bibnamefont {Gil-Narvion}}, \bibinfo
  {author} {\bibfnamefont {A.}~\bibnamefont {Gordillo-Guerrero}}, \bibinfo
  {author} {\bibfnamefont {D.}~\bibnamefont {Iniguez}}, \bibinfo {author}
  {\bibfnamefont {A.}~\bibnamefont {Maiorano}}, \bibinfo {author}
  {\bibfnamefont {M.}~\bibnamefont {F.}}, \bibinfo {author} {\bibfnamefont
  {E.}~\bibnamefont {Marinari}}, \bibinfo {author} {\bibfnamefont
  {V.}~\bibnamefont {Martin-Mayor}}, \bibinfo {author} {\bibfnamefont
  {J.}~\bibnamefont {Monforte-Garcia}}, \bibinfo {author} {\bibfnamefont
  {A.}~\bibnamefont {Muñoz~Sudupe}}, \bibinfo {author} {\bibfnamefont
  {D.}~\bibnamefont {Navarro}}, \bibinfo {author} {\bibfnamefont
  {G.}~\bibnamefont {Parisi}}, \bibinfo {author} {\bibfnamefont
  {S.}~\bibnamefont {Perez-Gaviro}}, \bibinfo {author} {\bibfnamefont
  {M.}~\bibnamefont {Pivanti}}, \bibinfo {author} {\bibfnamefont
  {F.}~\bibnamefont {Ricci-Tersenghi}}, \bibinfo {author} {\bibfnamefont
  {J.~J.}\ \bibnamefont {Ruiz-Lorenzo}}, \bibinfo {author} {\bibfnamefont
  {S.~F.}\ \bibnamefont {Schifano}}, \bibinfo {author} {\bibfnamefont
  {B.}~\bibnamefont {Seoane}}, \bibinfo {author} {\bibfnamefont
  {A.}~\bibnamefont {Tarancon}}, \bibinfo {author} {\bibfnamefont
  {R.}~\bibnamefont {Tripiccione}}, \ and\ \bibinfo {author} {\bibfnamefont
  {D.}~\bibnamefont {Yllanes}},\ }\href@noop {} {\  (\bibinfo {year}
  {{2013}}{\natexlab{a}})},\ \Eprint {http://arxiv.org/abs/arXiv:1307.4998}
  {arXiv:1307.4998} \BibitemShut {NoStop}%
\bibitem [{\citenamefont {Hukushima}\ and\ \citenamefont
  {Nemoto}(1996)}]{hukushima:96}%
  \BibitemOpen
  \bibfield  {author} {\bibinfo {author} {\bibfnamefont {K.}~\bibnamefont
  {Hukushima}}\ and\ \bibinfo {author} {\bibfnamefont {K.}~\bibnamefont
  {Nemoto}},\ }\href {\doibase 10.1143/JPSJ.65.1604} {\bibfield  {journal}
  {\bibinfo  {journal} {J. Phys. Soc. Japan}\ }\textbf {\bibinfo {volume}
  {65}},\ \bibinfo {pages} {1604} (\bibinfo {year} {1996})},\ \Eprint
  {http://arxiv.org/abs/arXiv:cond-mat/9512035} {arXiv:cond-mat/9512035}
  \BibitemShut {NoStop}%
\bibitem [{\citenamefont {Marinari}(1998)}]{marinari:98b}%
  \BibitemOpen
  \bibfield  {author} {\bibinfo {author} {\bibfnamefont {E.}~\bibnamefont
  {Marinari}},\ }in\ \href@noop {} {\emph {\bibinfo {booktitle} {Advances in
  Computer Simulation}}},\ \bibinfo {editor} {edited by\ \bibinfo {editor}
  {\bibfnamefont {J.}~\bibnamefont {Kerst\'esz}}\ and\ \bibinfo {editor}
  {\bibfnamefont {I.}~\bibnamefont {Kondor}}}\ (\bibinfo  {publisher}
  {Springer-Berlag},\ \bibinfo {year} {1998})\BibitemShut {NoStop}%
\bibitem [{\citenamefont {Alvarez~Ba{\~n}os}\ \emph
  {et~al.}(2010{\natexlab{b}})\citenamefont {Alvarez~Ba{\~n}os}, \citenamefont
  {Cruz}, \citenamefont {Fernandez}, \citenamefont {Gil-Narvion}, \citenamefont
  {Gordillo-Guerrero}, \citenamefont {Guidetti}, \citenamefont {Maiorano},
  \citenamefont {Mantovani}, \citenamefont {Marinari}, \citenamefont
  {Martin-Mayor}, \citenamefont {Monforte-Garcia}, \citenamefont
  {Mu{\~n}oz~Sudupe}, \citenamefont {Navarro}, \citenamefont {Parisi},
  \citenamefont {Perez-Gaviro}, \citenamefont {Ruiz-Lorenzo}, \citenamefont
  {Schifano}, \citenamefont {Seoane}, \citenamefont {Tarancon}, \citenamefont
  {Tripiccione},\ and\ \citenamefont {Yllanes}}]{janus:10}%
  \BibitemOpen
  \bibfield  {author} {\bibinfo {author} {\bibfnamefont {R.}~\bibnamefont
  {Alvarez~Ba{\~n}os}}, \bibinfo {author} {\bibfnamefont {A.}~\bibnamefont
  {Cruz}}, \bibinfo {author} {\bibfnamefont {L.~A.}\ \bibnamefont {Fernandez}},
  \bibinfo {author} {\bibfnamefont {J.~M.}\ \bibnamefont {Gil-Narvion}},
  \bibinfo {author} {\bibfnamefont {A.}~\bibnamefont {Gordillo-Guerrero}},
  \bibinfo {author} {\bibfnamefont {M.}~\bibnamefont {Guidetti}}, \bibinfo
  {author} {\bibfnamefont {A.}~\bibnamefont {Maiorano}}, \bibinfo {author}
  {\bibfnamefont {F.}~\bibnamefont {Mantovani}}, \bibinfo {author}
  {\bibfnamefont {E.}~\bibnamefont {Marinari}}, \bibinfo {author}
  {\bibfnamefont {V.}~\bibnamefont {Martin-Mayor}}, \bibinfo {author}
  {\bibfnamefont {J.}~\bibnamefont {Monforte-Garcia}}, \bibinfo {author}
  {\bibfnamefont {A.}~\bibnamefont {Mu{\~n}oz~Sudupe}}, \bibinfo {author}
  {\bibfnamefont {D.}~\bibnamefont {Navarro}}, \bibinfo {author} {\bibfnamefont
  {G.}~\bibnamefont {Parisi}}, \bibinfo {author} {\bibfnamefont
  {S.}~\bibnamefont {Perez-Gaviro}}, \bibinfo {author} {\bibfnamefont {J.~J.}\
  \bibnamefont {Ruiz-Lorenzo}}, \bibinfo {author} {\bibfnamefont {S.~F.}\
  \bibnamefont {Schifano}}, \bibinfo {author} {\bibfnamefont {B.}~\bibnamefont
  {Seoane}}, \bibinfo {author} {\bibfnamefont {A.}~\bibnamefont {Tarancon}},
  \bibinfo {author} {\bibfnamefont {R.}~\bibnamefont {Tripiccione}}, \ and\
  \bibinfo {author} {\bibfnamefont {D.}~\bibnamefont {Yllanes}} (\bibinfo
  {collaboration} {Janus Collaboration}),\ }\href {\doibase
  10.1088/1742-5468/2010/06/P06026} {\bibfield  {journal} {\bibinfo  {journal}
  {J. Stat. Mech.}\ ,\ \bibinfo {pages} {P06026}} (\bibinfo {year}
  {2010}{\natexlab{b}})},\ \Eprint {http://arxiv.org/abs/arXiv:1003.2569}
  {arXiv:1003.2569} \BibitemShut {NoStop}%
\bibitem [{\citenamefont {Fernandez}\ \emph {et~al.}(2008)\citenamefont
  {Fernandez}, \citenamefont {Maiorano}, \citenamefont {Marinari},
  \citenamefont {Martin-Mayor}, \citenamefont {Navarro}, \citenamefont
  {Sciretti}, \citenamefont {Tarancon},\ and\ \citenamefont
  {Velasco}}]{fernandez:08b}%
  \BibitemOpen
  \bibfield  {author} {\bibinfo {author} {\bibfnamefont {L.~A.}\ \bibnamefont
  {Fernandez}}, \bibinfo {author} {\bibfnamefont {A.}~\bibnamefont {Maiorano}},
  \bibinfo {author} {\bibfnamefont {E.}~\bibnamefont {Marinari}}, \bibinfo
  {author} {\bibfnamefont {V.}~\bibnamefont {Martin-Mayor}}, \bibinfo {author}
  {\bibfnamefont {D.}~\bibnamefont {Navarro}}, \bibinfo {author} {\bibfnamefont
  {D.}~\bibnamefont {Sciretti}}, \bibinfo {author} {\bibfnamefont
  {A.}~\bibnamefont {Tarancon}}, \ and\ \bibinfo {author} {\bibfnamefont
  {J.~L.}\ \bibnamefont {Velasco}},\ }\href {\doibase
  10.1103/PhysRevB.77.104432} {\bibfield  {journal} {\bibinfo  {journal} {Phys.
  Rev. B}\ }\textbf {\bibinfo {volume} {77}},\ \bibinfo {pages} {104432}
  (\bibinfo {year} {2008})},\ \Eprint {http://arxiv.org/abs/arXiv:0710.4246}
  {arXiv:0710.4246} \BibitemShut {NoStop}%
\bibitem [{\citenamefont {Cooper}\ \emph {et~al.}(1982)\citenamefont {Cooper},
  \citenamefont {Freedman},\ and\ \citenamefont {Preston}}]{cooper:82}%
  \BibitemOpen
  \bibfield  {author} {\bibinfo {author} {\bibfnamefont {F.}~\bibnamefont
  {Cooper}}, \bibinfo {author} {\bibfnamefont {B.}~\bibnamefont {Freedman}}, \
  and\ \bibinfo {author} {\bibfnamefont {D.}~\bibnamefont {Preston}},\ }\href
  {\doibase 10.1016/0550-3213(82)90240-1} {\bibfield  {journal} {\bibinfo
  {journal} {Nucl. Phys. B}\ }\textbf {\bibinfo {volume} {210}},\ \bibinfo
  {pages} {210} (\bibinfo {year} {1982})}\BibitemShut {NoStop}%
\bibitem [{\citenamefont {Amit}\ and\ \citenamefont
  {Martin-Mayor}(2005)}]{amit:05}%
  \BibitemOpen
  \bibfield  {author} {\bibinfo {author} {\bibfnamefont {D.~J.}\ \bibnamefont
  {Amit}}\ and\ \bibinfo {author} {\bibfnamefont {V.}~\bibnamefont
  {Martin-Mayor}},\ }\href {\doibase 10.1142/9789812775313_bmatter} {\emph
  {\bibinfo {title} {Field Theory, the Renormalization Group and Critical
  Phenomena}}},\ \bibinfo {edition} {3rd}\ ed.\ (\bibinfo  {publisher} {World
  Scientific},\ \bibinfo {address} {Singapore},\ \bibinfo {year}
  {2005})\BibitemShut {NoStop}%
\bibitem [{\citenamefont {Billoire}\ \emph {et~al.}(2011)\citenamefont
  {Billoire}, \citenamefont {Fernandez}, \citenamefont {Maiorano},
  \citenamefont {Marinari}, \citenamefont {Martin-Mayor},\ and\ \citenamefont
  {Yllanes}}]{billoire:11}%
  \BibitemOpen
  \bibfield  {author} {\bibinfo {author} {\bibfnamefont {A.}~\bibnamefont
  {Billoire}}, \bibinfo {author} {\bibfnamefont {L.~A.}\ \bibnamefont
  {Fernandez}}, \bibinfo {author} {\bibfnamefont {A.}~\bibnamefont {Maiorano}},
  \bibinfo {author} {\bibfnamefont {E.}~\bibnamefont {Marinari}}, \bibinfo
  {author} {\bibfnamefont {V.}~\bibnamefont {Martin-Mayor}}, \ and\ \bibinfo
  {author} {\bibfnamefont {D.}~\bibnamefont {Yllanes}},\ }\href {\doibase
  10.1088/1742-5468/2011/10/P10019} {\bibfield  {journal} {\bibinfo  {journal}
  {J. Stat. Mech.}\ ,\ \bibinfo {pages} {P10019}} (\bibinfo {year} {2011})},\
  \Eprint {http://arxiv.org/abs/arXiv:1108.1336} {arXiv:1108.1336} \BibitemShut
  {NoStop}%
\bibitem [{\citenamefont {Nightingale}(1975)}]{nightingale:76}%
  \BibitemOpen
  \bibfield  {author} {\bibinfo {author} {\bibfnamefont {M.~P.}\ \bibnamefont
  {Nightingale}},\ }\href {\doibase 10.1016/0378-4371(75)90021-7} {\bibfield
  {journal} {\bibinfo  {journal} {Physica A}\ }\textbf {\bibinfo {volume}
  {83}},\ \bibinfo {pages} {561} (\bibinfo {year} {1975})}\BibitemShut
  {NoStop}%
\bibitem [{\citenamefont {Ballesteros}\ \emph {et~al.}(1996)\citenamefont
  {Ballesteros}, \citenamefont {Fernandez}, \citenamefont {Martin-Mayor},\ and\
  \citenamefont {Mu{\~n}oz~Sudupe}}]{ballesteros:96}%
  \BibitemOpen
  \bibfield  {author} {\bibinfo {author} {\bibfnamefont {H.~G.}\ \bibnamefont
  {Ballesteros}}, \bibinfo {author} {\bibfnamefont {L.~A.}\ \bibnamefont
  {Fernandez}}, \bibinfo {author} {\bibfnamefont {V.}~\bibnamefont
  {Martin-Mayor}}, \ and\ \bibinfo {author} {\bibfnamefont {A.}~\bibnamefont
  {Mu{\~n}oz~Sudupe}},\ }\href {\doibase 10.1016/0370-2693(96)00358-9}
  {\bibfield  {journal} {\bibinfo  {journal} {Phys. Lett. B}\ }\textbf
  {\bibinfo {volume} {378}},\ \bibinfo {pages} {207} (\bibinfo {year}
  {1996})},\ \Eprint {http://arxiv.org/abs/arXiv:hep-lat/9511003}
  {arXiv:hep-lat/9511003} \BibitemShut {NoStop}%
\bibitem [{\citenamefont {Ballesteros}\ \emph {et~al.}(1998)\citenamefont
  {Ballesteros}, \citenamefont {Fernandez}, \citenamefont {Martin-Mayor},
  \citenamefont {Mu{\~n}oz~Sudupe}, \citenamefont {Parisi},\ and\ \citenamefont
  {Ruiz-Lorenzo}}]{ballesteros:98b}%
  \BibitemOpen
  \bibfield  {author} {\bibinfo {author} {\bibfnamefont {H.~G.}\ \bibnamefont
  {Ballesteros}}, \bibinfo {author} {\bibfnamefont {L.~A.}\ \bibnamefont
  {Fernandez}}, \bibinfo {author} {\bibfnamefont {V.}~\bibnamefont
  {Martin-Mayor}}, \bibinfo {author} {\bibfnamefont {A.}~\bibnamefont
  {Mu{\~n}oz~Sudupe}}, \bibinfo {author} {\bibfnamefont {G.}~\bibnamefont
  {Parisi}}, \ and\ \bibinfo {author} {\bibfnamefont {J.~J.}\ \bibnamefont
  {Ruiz-Lorenzo}},\ }\href {\doibase 10.1103/PhysRevB.58.2740} {\bibfield
  {journal} {\bibinfo  {journal} {Phys. Rev. B}\ }\textbf {\bibinfo {volume}
  {58}},\ \bibinfo {pages} {2740} (\bibinfo {year} {1998})}\BibitemShut
  {NoStop}%
\bibitem [{\citenamefont {Campos}\ \emph {et~al.}(2006)\citenamefont {Campos},
  \citenamefont {Cotallo-Aban}, \citenamefont {Martin-Mayor}, \citenamefont
  {Perez-Gaviro},\ and\ \citenamefont {Tarancon}}]{campos:06}%
  \BibitemOpen
  \bibfield  {author} {\bibinfo {author} {\bibfnamefont {I.}~\bibnamefont
  {Campos}}, \bibinfo {author} {\bibfnamefont {M.}~\bibnamefont
  {Cotallo-Aban}}, \bibinfo {author} {\bibfnamefont {V.}~\bibnamefont
  {Martin-Mayor}}, \bibinfo {author} {\bibfnamefont {S.}~\bibnamefont
  {Perez-Gaviro}}, \ and\ \bibinfo {author} {\bibfnamefont {A.}~\bibnamefont
  {Tarancon}},\ }\href {\doibase 10.1103/PhysRevLett.97.217204} {\bibfield
  {journal} {\bibinfo  {journal} {Phys. Rev. Lett.}\ }\textbf {\bibinfo
  {volume} {97}},\ \bibinfo {pages} {217204} (\bibinfo {year}
  {2006})}\BibitemShut {NoStop}%
\bibitem [{\citenamefont {Fernandez}\ \emph {et~al.}(2009)\citenamefont
  {Fernandez}, \citenamefont {Martin-Mayor}, \citenamefont {Perez-Gaviro},
  \citenamefont {Tarancon},\ and\ \citenamefont {Young}}]{fernandez:09b}%
  \BibitemOpen
  \bibfield  {author} {\bibinfo {author} {\bibfnamefont {L.~A.}\ \bibnamefont
  {Fernandez}}, \bibinfo {author} {\bibfnamefont {V.}~\bibnamefont
  {Martin-Mayor}}, \bibinfo {author} {\bibfnamefont {S.}~\bibnamefont
  {Perez-Gaviro}}, \bibinfo {author} {\bibfnamefont {A.}~\bibnamefont
  {Tarancon}}, \ and\ \bibinfo {author} {\bibfnamefont {A.~P.}\ \bibnamefont
  {Young}},\ }\href {\doibase 10.1103/PhysRevB.80.024422} {\bibfield  {journal}
  {\bibinfo  {journal} {Phys. Rev. B}\ }\textbf {\bibinfo {volume} {80}},\
  \bibinfo {pages} {024422} (\bibinfo {year} {2009})}\BibitemShut {NoStop}%
\bibitem [{\citenamefont {Leuzzi}\ \emph {et~al.}(2008)\citenamefont {Leuzzi},
  \citenamefont {Parisi}, \citenamefont {Ricci-Tersenghi},\ and\ \citenamefont
  {Ruiz-Lorenzo}}]{leuzzi:08}%
  \BibitemOpen
  \bibfield  {author} {\bibinfo {author} {\bibfnamefont {L.}~\bibnamefont
  {Leuzzi}}, \bibinfo {author} {\bibfnamefont {G.}~\bibnamefont {Parisi}},
  \bibinfo {author} {\bibfnamefont {F.}~\bibnamefont {Ricci-Tersenghi}}, \ and\
  \bibinfo {author} {\bibfnamefont {J.~J.}\ \bibnamefont {Ruiz-Lorenzo}},\
  }\href {\doibase 10.1103/PhysRevLett.101.107203} {\bibfield  {journal}
  {\bibinfo  {journal} {Phys. Rev. Lett.}\ }\textbf {\bibinfo {volume} {101}},\
  \bibinfo {pages} {107203} (\bibinfo {year} {2008})}\BibitemShut {NoStop}%
\bibitem [{\citenamefont {Ba\~{n}os}\ \emph
  {et~al.}(2012{\natexlab{b}})\citenamefont {Ba\~{n}os}, \citenamefont
  {Fernandez}, \citenamefont {Martin-Mayor},\ and\ \citenamefont
  {Young}}]{banos:12}%
  \BibitemOpen
  \bibfield  {author} {\bibinfo {author} {\bibfnamefont {R.~A.}\ \bibnamefont
  {Ba\~{n}os}}, \bibinfo {author} {\bibfnamefont {L.~A.}\ \bibnamefont
  {Fernandez}}, \bibinfo {author} {\bibfnamefont {V.}~\bibnamefont
  {Martin-Mayor}}, \ and\ \bibinfo {author} {\bibfnamefont {A.~P.}\
  \bibnamefont {Young}},\ }\href {\doibase 10.1103/PhysRevB.86.134416}
  {\bibfield  {journal} {\bibinfo  {journal} {Phys. Rev. B}\ }\textbf {\bibinfo
  {volume} {86}},\ \bibinfo {pages} {134416} (\bibinfo {year}
  {2012}{\natexlab{b}})},\ \Eprint {http://arxiv.org/abs/arXiv:1207.7014}
  {arXiv:1207.7014} \BibitemShut {NoStop}%
\bibitem [{\citenamefont {Baity-Jesi}\ \emph
  {et~al.}(2013{\natexlab{b}})\citenamefont {Baity-Jesi}, \citenamefont
  {Fernandez}, \citenamefont {Martin-Mayor},\ and\ \citenamefont
  {Sanz}}]{baityjesi:13}%
  \BibitemOpen
  \bibfield  {author} {\bibinfo {author} {\bibfnamefont {M.}~\bibnamefont
  {Baity-Jesi}}, \bibinfo {author} {\bibfnamefont {L.~A.}\ \bibnamefont
  {Fernandez}}, \bibinfo {author} {\bibfnamefont {V.}~\bibnamefont
  {Martin-Mayor}}, \ and\ \bibinfo {author} {\bibfnamefont {J.~M.}\
  \bibnamefont {Sanz}},\ }\href@noop {} {\  (\bibinfo {year}
  {2013}{\natexlab{b}})},\ \Eprint {http://arxiv.org/abs/arXiv:1309.1599}
  {arXiv:1309.1599} \BibitemShut {NoStop}%
\bibitem [{\citenamefont {Fernandez}\ \emph {et~al.}(2011)\citenamefont
  {Fernandez}, \citenamefont {Martin-Mayor},\ and\ \citenamefont
  {Yllanes}}]{fernandez:11b}%
  \BibitemOpen
  \bibfield  {author} {\bibinfo {author} {\bibfnamefont {L.~A.}\ \bibnamefont
  {Fernandez}}, \bibinfo {author} {\bibfnamefont {V.}~\bibnamefont
  {Martin-Mayor}}, \ and\ \bibinfo {author} {\bibfnamefont {D.}~\bibnamefont
  {Yllanes}},\ }\href {\doibase 10.1103/PhysRevB.84.100408} {\bibfield
  {journal} {\bibinfo  {journal} {Phys. Rev. B}\ }\textbf {\bibinfo {volume}
  {84}},\ \bibinfo {pages} {100408(R)} (\bibinfo {year} {2011})},\ \Eprint
  {http://arxiv.org/abs/arXiv:1106.1555} {arXiv:1106.1555} \BibitemShut
  {NoStop}%
\bibitem [{\citenamefont {Fernandez}\ \emph {et~al.}(2012)\citenamefont
  {Fernandez}, \citenamefont {Gordillo-Guerrero}, \citenamefont
  {Martin-Mayor},\ and\ \citenamefont {Ruiz-Lorenzo}}]{fernandez:12}%
  \BibitemOpen
  \bibfield  {author} {\bibinfo {author} {\bibfnamefont {L.~A.}\ \bibnamefont
  {Fernandez}}, \bibinfo {author} {\bibfnamefont {A.}~\bibnamefont
  {Gordillo-Guerrero}}, \bibinfo {author} {\bibfnamefont {V.}~\bibnamefont
  {Martin-Mayor}}, \ and\ \bibinfo {author} {\bibfnamefont {J.~J.}\
  \bibnamefont {Ruiz-Lorenzo}},\ }\href {\doibase 10.1103/PhysRevB.86.184428}
  {\bibfield  {journal} {\bibinfo  {journal} {Phys. Rev. B}\ }\textbf {\bibinfo
  {volume} {86}},\ \bibinfo {pages} {184428} (\bibinfo {year} {2012})},\
  \Eprint {http://arxiv.org/abs/arXiv:1205.0247} {arXiv:1205.0247} \BibitemShut
  {NoStop}%
\bibitem [{\citenamefont {Fytas}\ and\ \citenamefont
  {Martin-Mayor}(2013)}]{fytas:13}%
  \BibitemOpen
  \bibfield  {author} {\bibinfo {author} {\bibfnamefont {N.~G.}\ \bibnamefont
  {Fytas}}\ and\ \bibinfo {author} {\bibfnamefont {V.}~\bibnamefont
  {Martin-Mayor}},\ }\href {\doibase 10.1103/PhysRevLett.110.227201} {\bibfield
   {journal} {\bibinfo  {journal} {Phys. Rev. Lett.}\ }\textbf {\bibinfo
  {volume} {110}},\ \bibinfo {pages} {227201} (\bibinfo {year} {2013})},\
  \Eprint {http://arxiv.org/abs/arXiv:1304.0318} {arXiv:1304.0318} \BibitemShut
  {NoStop}%
\bibitem [{Note1()}]{Note1}%
  \BibitemOpen
  \bibinfo {note} {Dimensionless quantities behave as a function of $L$ and the
  reduced temperature $t=(T-T_\protect \mathrm {c})/T_\protect \mathrm {c}$ as
  $$g(L,t)=f_g(L^{1/\nu } t) + L^{-\omega } h_g(L^{1/\nu }t)+\protect \ldots
  \protect \tmspace +\thinmuskip {.1667em},$$ where $f_g$ and $h_g$ are very
  smooth (actually analytical) scaling functions. In particular, we name $f_\xi
  $ and $g_\xi $ the scaling functions corresponding to $\xi /L$. In fact, see
  e.g. Ref.~\protect \rev@citealpnum {amit:05}, the detailed form of
  Eq.~\protect \textup {\hbox {\mathsurround \z@ \protect \normalfont
  (\ignorespaces \ref {eq:Tc-shift}\unskip \@@italiccorr )}} follows from the
  Taylor expansions $f_\xi (x)=f_\xi (0)+x f'_\xi (0)+\protect \ldots $ and
  $h_\xi (x)=h_\xi (0)+\protect \ldots $: $$ t^{\protect \tmspace +\thinmuskip
  {.1667em}\protect \mathrm {cross}}_{L,2L}=\protect \frac {h_\xi (0)}{f'_\xi
  (0)}\protect \tmspace +\thinmuskip {.1667em} \protect \frac {1-2^{-\omega
  }}{2^{1/\nu }-1}\protect \tmspace +\thinmuskip {.1667em} L^{-\omega -\protect
  \frac {1}{\nu }}\ +\ \protect \ldots \protect \tmspace +\thinmuskip
  {.1667em}. $$ A similar computation yields the amplitudes for the scaling
  corrections of $g_{2L}^\protect \mathrm {cross}$ and $g_L^\protect \mathrm
  {cross}$ in Eq.~\protect \textup {\hbox {\mathsurround \z@ \protect
  \normalfont (\ignorespaces \ref {eq:QO}\unskip \@@italiccorr )}}:
  $$A_g^{(2L)}=2^{1/\nu }\protect \frac {1-2^{-\omega }}{2^{1/\nu }-1}h_\xi
  (0)\protect \frac {f_g'(0)}{f_\xi '(0)}\ +\ 2^{-\omega } h_g(0)\protect
  \tmspace +\thinmuskip {.1667em},$$ and $$A_g^{(L)}=\protect \frac
  {1-2^{-\omega }}{2^{1/\nu }-1}h_\xi (0)\protect \frac {f_g'(0)}{f_\xi '(0)}\
  +\ h_g(0)\protect \tmspace +\thinmuskip {.1667em}$$ Either of the two
  amplitudes $A_g^{(L)}$, $A_g^{(2L)}$ can dominate, depending both on $g$ and
  on the magnitude chosen to find the crossing point ($\xi /L$, $U_4$,
  etc.)}\BibitemShut {NoStop}%
\bibitem [{\citenamefont {Yllanes}(2011)}]{yllanes:11}%
  \BibitemOpen
  \bibfield  {author} {\bibinfo {author} {\bibfnamefont {D.}~\bibnamefont
  {Yllanes}},\ }\href@noop {} {\emph {\bibinfo {title} {Rugged Free-Energy
  Landscapes in Disordered Spin Systems}}}\ (\bibinfo  {publisher} {Ph.D.
  thesis, UCM},\ \bibinfo {year} {2011})\ \Eprint
  {http://arxiv.org/abs/arXiv:1111.0266} {arXiv:1111.0266} \BibitemShut
  {NoStop}%
\bibitem [{Note2()}]{Note2}%
  \BibitemOpen
  \bibinfo {note} {As usual, $\beta =1/T$. We employ it in order to allow for a
  direct comparison with raw data from Ref.~\protect \rev@citealpnum
  {hasenbusch:08b}.}\BibitemShut {Stop}%
\end{thebibliography}
\end{document}